\documentclass[14pt, preprint2]{aastex62}
\graphicspath{{./}{Plots/}}
\usepackage{float}
\makeatletter
\let\newfloat\newfloat@ltx
\makeatother
\usepackage{graphicx, subfigure, algorithm, algorithmic, amsmath}
\newcommand{\project}[1]{\textsl{#1}}
 
\newcommand{\tc}{\project{The~Cannon}} 
\newcommand{\apogee}{\project{\textsc{apogee}}}

\newcommand{\Gaia}{\project{Gaia}}

\newcommand{\teff}{\mbox{$T_{\rm eff}$}}

\newcommand{\feh}{\mbox{$\rm [Fe/H]$}}

\newcommand{\alphafe}{\mbox{$\rm [\alpha/Fe]$}}

\newcommand{\logg}{\mbox{$\log g$}}

\newcommand{\rgal}{\mbox{$R_{\text{GAL}}$}}

\newcommand{\num}{27,135}

\begin{document}

\title{Tracing the assembly of the Milky Way's disk through abundance clustering }

\author{Bridget L. Ratcliffe}
\affil{Department of Statistics, Columbia University, 1255 Amsterdam Avenue, New York, NY 10027, USA}
\author{Melissa K. Ness}
\affil{Department of Astronomy, Columbia University, 550 West 120th Street, New York, NY, 10027, USA}
\affil{Center for Computational Astrophysics, Flatiron Institute, 162 Fifth Avenue, New York, NY, 10010, USA}
\author{Kathryn V. Johnston}
\affil{Department of Astronomy, Columbia University, 550 West 120th Street, New York, NY, 10027, USA}
\affil{Center for Computational Astrophysics, Flatiron Institute, 162 Fifth Avenue, New York, NY, 10010, USA}
\author{Bodhisattva Sen}
\affil{Department of Statistics, Columbia University, 1255 Amsterdam Avenue, New York, NY 10027, USA}

\begin{abstract}
A major goal in the field of galaxy formation is to understand the formation of the Milky Way's disk. 
The first step toward doing this is to empirically describe its present state. We use the new high-dimensional dataset of 19 abundances from  \num\ red clump \apogee\ stars to examine the distribution of clusters defined using abundances.  We  explore different dimensionality reduction techniques and implement a non-parametric agglomerate hierarchical clustering method. We see that groups defined using abundances are spatially separated, as a function of age. Furthermore, the abundance groups represent
different distributions in the [Fe/H]-age plane. Ordering our clusters by age reveals patterns suggestive of the sequence of chemical enrichment in the disk over time.
Our results indicate that a promising avenue to trace the details of the disk's assembly is via a full interpretation of the empirical connections we report.   

\end{abstract}

\section{Introduction} \label{sec:intro}

The Milky Way's stellar mass is concentrated in the disk, the chemodynamical characteristics of which can constrain the Galaxy's formation history. The spatial and kinematic distribution of the Milky Way's disk has historically been separated into two populations, termed the `thin' and `thick' disk \citep{gilmore2002deciphering}. 
These two populations have also been found to have different chemical properties, with the `thin' and `thick' disk structures having lower/higher [Fe/H] and higher/lower  [$\alpha$/Fe] respectively   \citep[e.g.][]{fuhrmann1998nearby, bensby2014exploring}. 
However, the disk's distribution has also been argued to be better represented by a set of populations with a continuous sequence in morphological parameters, rather than described by a strict binary division \citep{bovy2012milky}. 
 
The advent of large spectroscopic surveys has provided the opportunity to examine the disk's chemodynamical structure in new ways. The \apogee\ survey, in particular \cite{Majewski2017}, has afforded important insight into the disk, due to its deep infrared coverage \citep{zasowski2014mapping}, medium resolution spectra (R=22,500) \citep{Holtzman2015, GP2015} and large number of stars ($>$ 300,000 in DR14) \citep{Majewski2017,blanton2017sloan,abolfathi2018fourteenth}. Studies can now traverse a larger spatial extent \citep[e.g.][]{Nidever2014, Hayden2015} and simultaneously, across a diversity of chemical elements, beyond a bulk [Fe/H] and [$\alpha$/Fe] \citep[e.g.][]{Weinberg2019}.

\apogee\ revealed bimodality in the two-dimensional [Fe/H]-[$\alpha$/Fe] plane, with the stars clearly divided into a `high-$\alpha$' and `low-$\alpha$' sequence. These two sequences, of high- and low-$\alpha$ stars,  have different ages and spatial and kinematic distributions. The relative fraction of high- and low-$\alpha$ stars, respectively,  varies depending on disk height and radius, with old high-$\alpha$ stars living primarily in $0.5<|z|<2$ kpc and $3<\rgal\ <11$ kpc and younger low-$\alpha$ stars living in $|z| < 1$ kpc for $3<\rgal\ <9$ kpc and evenly spread throughout $|z|$ for \rgal\ $>9$ kpc \citep[e.g.][]{hayden2015chemical, nidever2014tracing, VSA2018}. Using stars that are sampled by \apogee\ and \Gaia, the high- and low-$\alpha$ sequences are seen to exhibit different orbital properties as well as age-velocity dispersion relations and distributions \citep[e.g.][]{mackereth2019dynamical}, including at fixed age \citep{Gandhi2019, blancato2019variations}.

It is tempting to associate these two sequences defined chemically directly with the `thin' and `thick' disks that have been defined in the stellar spatial and kinematic distribution, in apparent confirmation of the bimodal picture. 
However,  it has become clear that the spatial-to-chemical mapping is discordant across large spatial extents, with the geometry of the two disks defined spatially differing from those where the separation is made in the chemical plane  \citep{BH2019}.

The observed spatial and kinematic differences between the high- and low-$\alpha$ stars have led to the interpretation that there are two discrete modes of chemical evolution in the disk, with much debate regarding the origin of the sequences. Simulation work (\cite{mackereth2018origin}, \cite{Clarke2019}) has been pursued to find the sources of these differences. From 133 Milky Way-like EAGLE simulations, \cite{mackereth2018origin} find that the Milky Way bimodality in chemistry is rare, evolving in only 5\% of the simulation galaxies. With their simulations, they discuss how the bimodality is a consequence of an early gas accretion episode. On the other hand, \cite{Clarke2019} demonstrate with a GASOLINE simulation that the bimodality is natural if the early gas-rich disk fragments, causing a mixture of clumpy and distributed star formation. The clumps enrich rapidly and migrate from the low- to high-$\alpha$ sequence due to the clumps high star formation rate, while the distributed star formation produces the low-$\alpha$ sequence. 

Many different methods for dividing the stars in the [$\alpha$/Fe]-[Fe/H] plane have been used in order to aid with interpretation and comparison to simulations  (e.g. \cite{Masseron2015}, \cite{Weinberg2018}, \cite{blancato2019variations}). In addition to density contours, \citet{Masseron2015} used the [C/N] ratio to confirm their high- and low-$\alpha$ sequence populations. \cite{Weinberg2018} trace the groups in a two-dimensional density distribution and follow a shallow valley separating the two sequences in their division. \cite{blancato2019variations} employ a more robust separation through a soft clustering using a Gaussian mixture model in the [$\alpha$/Fe]-[Fe/H] plane, but found that their results were sensitive to the initialization parameters.

There are several limitations to these simple divisions in [$\alpha$/Fe]-[Fe/H].   There is no standard way to separate the stars into the two sequences, and many approaches rely on an ad-hoc division-by-eye. 
Moreover, any chemo-dynamical analysis of the disk that uses only [$\alpha$/Fe]-[Fe/H] potentially omits important features from the larger chemical space. Perceiving two clusters in the two-dimensional plane does not necessarily mean that there are only two populations in the larger abundance space, and this approach may effectively marginalize over potentially physically meaningful sub-structures or signatures. 

Large surveys that measure many dimensions of chemical-space, like \apogee, offer the opportunity to investigate abundance distributions more generally.

Working with high-dimensional data is nontrivial as the dimensions are difficult to visualize, there is a possibility of models over-fitting data, and the higher dimensional space will be sparse without a large sample size --- all often referred to as the curse of dimensionality. Previous work has attempted to reduce the high-dimensional data into something more manageable. \cite{ting2012principal} use Principal Component Analysis on 25 elements from HERMES in order to find the most informative lower dimensional space for high and low metallicity stars. They establish a connection between the principal components and nucleosythesis mechanisms, and also elude to the fact that too many redundant dimensions reduces our ability to locate clusters. Through the employment of Expectation Maximized Principal Component Analysis on the full spectra, \cite{PJ2018} find that only about 10 principal components accurately model the spectra. Recently, \cite{casey2019data} study a data-driven model of nucleosynthesis with chemical tagging in a lower-dimensional latent space on the GALAH dataset. They find $\sim$ 2,500 stars divide into three clusters in the latent chemical abundance space; however results were not consistent as more stars were added, admitting that a parametric Gaussian model is likely incorrect.  

In this paper, we seek to leverage the $>$ 20 element abundances measured by the \apogee\ survey to effectively organize stars by their overall chemical similarity.

Our approach is non-parametric and impartial to how many underlying populations comprise the data --- hierarchical clustering (\cite{ward1963hierarchical}) in a 19 dimensional chemical abundance space. We choose this method rather than something like K-means (e.g. \cite{Hogg2016}), as with hierarchical clustering there is no prior concern of choosing the correct number of clusters. Without being able to visualize all dimensions at once, choosing the wrong number of clusters could give rise to misleading results for a method requiring the number of clusters beforehand. Additionally, we investigate what abundances are driving the clustering results, and employ standard data reduction techniques Principal Component Analysis (\cite{pearson1901liii}, \cite{hotelling1933analysis}) and Isomap (\cite{tenenbaum2000global}) to determine the most important features of the data and ensure clustering results are not over-fit. We also consider how many groups are justified as discrete populations. 

This paper is organized as follows. In Section \ref{sec:data} we discuss the datasets used. Methods used for clustering, dimension reduction, and determining the number of stellar populations are outlined in Section \ref{sec:methods}. Section \ref{sec:results} briefly explores the abundance correlations, and investigates how the clusters defined from the large chemical abundance space appear in the [$\alpha$/Fe]-[Fe/H]-plane. In Section \ref{sec:groups}, the question of how many significantly different clusters is investigated. Finally, Sections \ref{sec:discussion} and \ref{sec:conclusions} present the main conclusions and discussion of our analysis.

\section{Data}
\label{sec:data}

In this paper, we use \apogee\ DR14 data (\cite{Majewski2017}, \cite{abolfathi2018fourteenth}) from SDSS-IV (\cite{blanton2017sloan}). We focus on the red clump stars found in the DR14 \apogee\ red clump catalogue (\cite{bovy2014apogee}), with their abundances processed by \apogee\ Stellar Parameter and Chemical Abundance Pipeline (ASPCAP) \citep{perez2016aspcap} and  ages calculated by \citet{Sanders2018}.
Stellar paramaters of \teff, \logg, and \feh\ as well as over 20 chemical element abundances are derived from the spectra with the ASPCAP pipeline.

\subsection{\apogee\ Red Clump Abundances and Ages}

In our analysis, we examine chemical abundance ratios for the red clump (RC) stars given in the DR14 \apogee\ red clump catalog. RC stars span a narrow parameter space in effective temperature, \teff, and surface gravity, \logg. This avoids systematic effects induced in abundances by astrophysical diffusion (\cite{Liu2019}) or imperfect stellar modeling (\cite{Weinberg2019}), creating a high fidelity sample of stars. As detailed in \cite{bovy2014apogee}, the RC catalogue was created through a combination of selection cuts in surface gravity, \teff, metallicity, and dereddened color $[J-K_s]_0$. Contamination from Red Giant Branch stars estimates to be less than 5\%. 

We also consider the stars' associated ages and errors calculated by \cite{Sanders2018}. Through a Bayesian artificial neural network, posterior estimates for distances, masses, and ages are found from spectroscopic, photometric, and astrometric data. The priors assigned reflect the stars likely population membership, with age priors differing depending on if the star is likely located in the thin disk, thick disk, bulge, or stellar halo.

For our analysis, we select the 19 abundances from the available set of 25 in the \apogee\ catalogue which we deem to have the most reliable measurements. The set of abundances selected to work with consists of Ni, Co, Fe, Mn, Cr, V, Ti, TiII, Ca, Si, Mg, O, K, S, P, Al, N, CI, and C. 

\subsection{Quality cuts on the data}
\label{sec:quality_cuts}

In order to remove outliers with anomalous abundance measurements, we consider stars with abundances -1 $\leq$ [X/Fe] $\leq$ 1 and -1 $\leq$ [Fe/H] $\leq$ 1. Similarly, since all but a handful of stars have ages greater than 0.1 Gyr, we remove the few very young stars. This provides a final sample of 27,135 RC stars with no missing parameters.

After removing outliers, the abundances still cover different ranges. For instance [Co/Fe] spans nearly the entire interval [-1, 1] while [Ni/Fe] values only fall between [-0.34, 0.39]. This variance in the range leads to biases in dimensionality reduction techniques such as Principal Component Analysis (PCA), which we employ in this paper. In order to combat this, we standardize the data to give each abundance mean 0 and standard deviation 1. From here on we refer to these standardized abundances simply as abundances.

\section{Methods}
\label{sec:methods}

In this section, we describe the methods used in our analysis. We first wish to examine how the data is organized into groups using a clustering algorithm. Further, how do the groups obtained with a clustering approach compare to the high- and low-$\alpha$ sequences as they are typically defined? The implementation of our selected clustering method, hierarchical clustering, is described in Section \ref{sec:Hierarchical_results}. In Section \ref{sec:lowerDim}, we compare the clustering performed in 19 dimensions to groups determined using lower dimensional representations of the data produced by PCA and Isomap. These  best capture the most important features of the data and are a test of how robust our results are when done using the full dimensionality of the available data. We also address the question of how many stellar populations our dataset can be decomposed into with clustering, and what this means. We examine how many groups justifiably exist in the abundance space using the Gap Statistic metric (Section \ref{sec:groups}) and Dunnett's modification of the Tukey-Kramer method (Section \ref{sec:groups_age}).

\subsection{Clustering Methodologies} 

\subsubsection{Agglomerative Hierarchical Clustering}
\label{sec:ward}

Agglomerative hierarchical clustering is a clustering method that organizes stars into groups (\cite{jain1988algorithms}), working from individual data points (in our case each star with its set of abundances) that hierarchically merge into groups (in our case defined by their abundance similarity). The resulting clusters contain the stars that are most similar to one another, while the clusters themselves are most dissimilar from each other. The metric that splits the difference between what is similar and different is left up to the user. We discuss our choice of metric in Sections \ref{sec:Hierarchical_results} and \ref{sec:groups_age}. 

Many clustering methods, such as K-means (\cite{hartigan1979algorithm}), require prior knowledge for how many clusters comprise the data. Hierarchical clustering, however, does not require a predefined number of clusters to be set. Rather, it hierarchically builds groups from the number of data points, N, to one single group, in a tree like structure. That is, it first merges the most similar two stars and moves up the hierarchy until only one cluster remains after combining the least similar groups. The user can employ a metric to indicate the maximum number of different clusters that are justified given the data.  The output is expressed in a so-called dendrogram, or tree diagram. This dendrogram illustrates the hierarchical arrangement of the clusters and allows us to visualise the similarity structure of the data. A dendrogram  shows the transition from N $=27,135$ groups to 1, where all 27,135 stars are plotted as the leaves of the tree and groups combine at certain ``heights" to create nodes. In the metric we choose to implement for the clustering, the node heights represent the total within cluster sum of squares error for clusters $K=1,...,27,135$, $$\sum_{k=1}^K \sum_{i\in S_k}\sum_{j=1}^{19}(x_{ij}-\bar{x}_{kj})^2,$$ where $S_k$ is the set of stars in cluster $k$, $x_{ij}$ is the $j^\text{th}$ abundance of star $i$, and $\bar{x}_{kj}$ is the average of the $j^\text{th}$ abundance for $k^\text{th}$ cluster. While a larger node height indicates the combining groups are more dissimilar, the difference between subsequent node heights can be thought of as a potential; the larger the potential, the more confidently we can say that the two adjoining groups are distinct. Once the user views the structure of the dendrogram, they are able to decide an appropriate number of clusters (\cite{ward1963hierarchical}). Thus hierarchical clustering suits not only our goal to cluster the data into two sequences, but also to determine if the RC sample form more than two populations in the larger abundance space.

We choose to cluster the data via agglomerative hierarchical clustering, using Ward's Minimum Variance Criterion (\cite{ward1963hierarchical}), which minimizes the total within-cluster variance at each step. This dissimilarity measure maximizes the distances between clusters while minimizing the distances between stars within a cluster (\cite{ward1963hierarchical}). This measure aligns with our goals for classifying stellar populations, where stars most similar in chemical space should be grouped together. 

Specifically, we use the Ward2 algorithm described in \cite{kaufman2009finding} and \cite{murtagh2011ward}. Beginning with each star as its own cluster, at each step we combine the pair of clusters that leads to a minimum increase in total within-cluster variance until only one large cluster containing all the stars remains. The explicit steps are given as Algorithm \ref{alg:Wards} in the Appendix.

\subsubsection{Principal Component Analysis}

Principal Component Analysis (PCA; \cite{pearson1901liii}, \cite{hotelling1933analysis}) is a linear dimensionality reduction technique. In our case, this transforms the data from a 19 dimensional abundance space to a much smaller set of variables that contain the most information about the data. The new smaller linearly uncorrelated set of variables, or principal components, are linear combinations of the 19 abundances. The first of these components is in the direction of the largest spread in the data. The second component, third component and so forth, lie in the direction of the maximum variance subject to being uncorrelated to the previous components (\cite{jolliffe1990principal}, \cite{ringner2008principal}).

As discussed in Section \ref{sec:quality_cuts}, we first normalize, or standardize, each variable to have mean 0 and standard deviation 1. Before standardization, each variable spans a different range, and if used directly, this would prevent each abundance from having equal importance in the analysis. Instead,  a few variables would dominate and skew the results.  Therefore, the $19 \times 19$ abundance covariance matrix that is calculated for PCA is computed using the standardized abundances as inputs. The covariance between any two abundances, $x$ and $y$, is given as $$\frac{1}{n-1}\sum_{i=1}^n x_i y_i.$$ The eigenvectors and eigenvalues of this covariance matrix are then calculated. It is established that the largest eigenvalues contain the most useful information regarding the spread of the data, and that the smallest eigenvalues primarily capture the noise. In fact, the eigenvalues capture the amount of variance explained by the associated eigenvector (\cite{wold1987principal}). Therefore, the first principal component is the eigenvector corresponding to the largest eigenvalue and so on (\cite{jolliffe1990principal}). Additionally, the principal components are scaled to have unit norm in order to easily compare the amount each abundance contributes to a specific component.

\subsubsection{Isomap}
\label{sec:Isomap}

While PCA is beneficial as a linear dimensionality reduction methodology, a more generalized approach is to assume that our data lie on an embedded non-linear manifold within the 19 dimensional space. This assumption requires a more flexible model that is able to capture additional (non-linear) structure in the data, which may not be revealed with PCA. The nonlinear dimension reduction technique, Isomap (\cite{tenenbaum2000global}), finds a low-dimensional embedding of the data that preserves the geodesic distances from 19 dimensions. Here, the geodesic distances represent the distances between stars in a neighborhood graph, as explained below. With the advantage of being less sensitive to noise, Isomap more reliably represents the data's global structure in meaningful coordinates on a lower dimensional Euclidean space (\cite{rosman2010nonlinear}, \cite{silva2003global}, \cite{tenenbaum2000global}). By comparing the results to PCA, we hope to understand the key features of the data.

The algorithm is introduced in \cite{tenenbaum2000global} and contains three primary steps. It is detailed in Algorithm \ref{alg:Isomap} in the Appendix and described here. The first step establishes the 5 nearest neighbors for each data point. Star $j$ is a neighbor of star $i$ if it is one of the 5 closest stars to star $i$ in the 19 dimensional abundance space using a euclidean distance calculation. This creates a neighborhood graph $G$, where the edge length, $G_{i,j}$, is equal to the euclidean distance between the two neighbors if star $j$ is a neighbor of star $i$, and is set to 0 if not. 

The second step creates a geodesic distance matrix, $d_G$. This approximates the shortest graph distance between star $i$ and star $j$, using Dijkstra's algorithm (\cite{dijkstra1959note}), on the neighborhood graph $G$ for all stars, $i$ and $j$. Following the optimal path between nearest neighbors at a given stage, Dijkstra's algorithm finds the shortest path from star $i$ to star $j$, $d_G(i,j)$. The explicit steps of the algorithm are given in Algorithm \ref{alg:dijkstra}, in the Appendix. 

In the last step, Isomap applies classical Multidimensional Scaling (MDS; \cite{kruskal1964multidimensional}) to the geodesic distance matrix, $d_G$, found in the previous step. MDS solves the problem of creating a coordinate matrix from distances provided (\cite{kruskal1964multidimensional}) and is given in detail in the Appendix as Algorithm \ref{alg:mds}. For our analysis, we choose to project our data into two dimensions. 

\subsection{Metrics to Quantify Cluster Significance}

\subsubsection{The Gap Statistic}
\label{sec:gapDef}

Proposed by \cite{tibshirani2001estimating}, the gap statistic is a method for estimating the number of clusters in a dataset by formalizing the elbow method. The elbow method is explained in detail in \cite{kodinariya2013review}. Briefly, in a plot of total within-cluster sum of squares versus number of clusters, the elbow method requires one to infer where an ``elbow" lies. The elbow method is usually subjective since in practice it can be unclear where exactly to cut off the number of clusters.

Assuming $k$ clusters, in our analysis $W_k$ is defined as the pooled within-cluster sum of squares about the cluster means. The goal is to compare $\log(W_k)$ to what the expected value would be if the data were from a reference distribution, $E^*[\log(W_k)]$. The reference distribution is typically a uniform distribution taken over the range of observed values, with $E^*[\log(W_k)]$ being the mean log$(W_k)$ for many bootstrapped samples from the reference distribution. The gap statistic is then defined as $$Gap(k) = E^*[\log(W_k)]-\log(W_k)$$

The determined number of clusters $\hat{k}$ is the smallest value $k$ such that $Gap(k) \geq Gap(k+1)-s_{k+1}$ where $s_k$ is the standard deviation of $E^*[\log(W_k)]$ for $k$ clusters from Monte Carlo simulations.

\subsubsection{Dunnett's modification of the Tukey-Kramer method for determining cluster significance as a function of age}
\label{sec:ttest}

To find the maximum number of groups with statistically significant values that we do not cluster using --- in our case different mean ages in Section \ref{sec:groups_age} --- we use Dunnett's modification of the Tukey-Kramer method to compare the mean ages of the clusters. Due to our large stellar sample, we choose a pairwise multiple comparison test over the Kolmogorov-Smirnov test, (\cite{massey1951kolmogorov}) which compares distributions. 

For each cluster of stars, $k$, we can calculate a mean age $\mu_k$. We wish to evaluate if our clusters are representative of populations with significantly different mean ages. To test the statistical significance of the mean age differences, we conduct a pairwise multiple comparison test, which compares the means of clusters 1 and 2, clusters 1 and 3, clusters 2 and 3, and so on. To quantify our results, we choose the confidence level to be the standard value 0.95 to create confidence intervals. A confidence level of 1-$\alpha$ means that if many samples were drawn using a given true parameter, we would expect the true parameter to fall in (1-$\alpha$)$\times$100\% of the confidence intervals. 

We specifically implement Dunnett's modification of the Tukey-Kramer method (\cite{dunnett1980pairwise}), which allows for the clusters to have unequal sample sizes and different variances. In order to determine if two clusters $i$ and $j$ have significantly different mean ages, we look at the confidence interval $$\mu_i-\mu_j \pm A_{ij, \alpha, k}\Big(\frac{s_i^2}{n_i} + \frac{s_j^2}{n_j}\Big)^{1/2},$$ where $n_i, n_j$ are the number of stars in group $i,j$ respectively, and $s_i,s_j$ are the unbiased estimated standard deviations given by $$s_i = \sqrt{\frac{1}{n_i-1}\sum_{x=1}^{n_i}(a_x-\bar{a}_i)^2}$$ $$s_j = \sqrt{\frac{1}{n_j-1}\sum_{x=1}^{n_j}(a_x-\bar{a}_j)^2}$$ for ages $a$ with $\bar{a}_i$ and $\bar{a}_j$ being the mean ages for group $i$ and $j$. Let $q_{\alpha, k, \nu}$ be the critical value of Studentized range distribution --- with parameters $k$ clusters and degrees of freedom $\nu$ --- that gives the boundary of the acceptance region for the test with confidence level $1-\alpha$. Then, $A_{ij,\alpha,k}$ is a weighted average of the critical values given by  $$A_{ij,\alpha,k} = \frac{1}{\sqrt{2}}\frac{q_{\alpha,k, n_i-k}s_i^2/n_i + q_{\alpha,k, n_j-k}s_j^2/n_j}{s_i^2/n_i + s_j^2/n_j}$$

In order to determine the largest number of clusters with distinct age distributions, we start with two clusters and increase the number of groups until the mean ages are no longer significantly different. We begin by evaluating if $\mu_1$ is significantly different than $\mu_2$ when the data is split into two clusters. We claim the two cluster means are significantly different if the confidence interval does not contain 0, and show no significant difference if 0 is included. If 0 is not contained in the interval, we then split the data into three clusters and compare $\mu_1$ to $\mu_2$, $\mu_1$ to $\mu_3$, and $\mu_2$ to $\mu_3$. If every one of these tests returns the means are significantly different then we proceed onto four clusters and compare $\mu_1$ to $\mu_2$, $\mu_1$ to $\mu_3$, $\mu_1$ to $\mu_4$ and so on. We continue to increase the number of clusters until one of the pairwise tests concludes that at least two means are not significantly different. Say this happens when there are $K^*+1$ clusters. Then the maximum number of clusters with distinct ages is $K^*$.

\section{Results I: Clusters in abundance space projections} 
\label{sec:results}

\begin{figure}[]
     \centering
     \includegraphics[width=.4\textwidth]{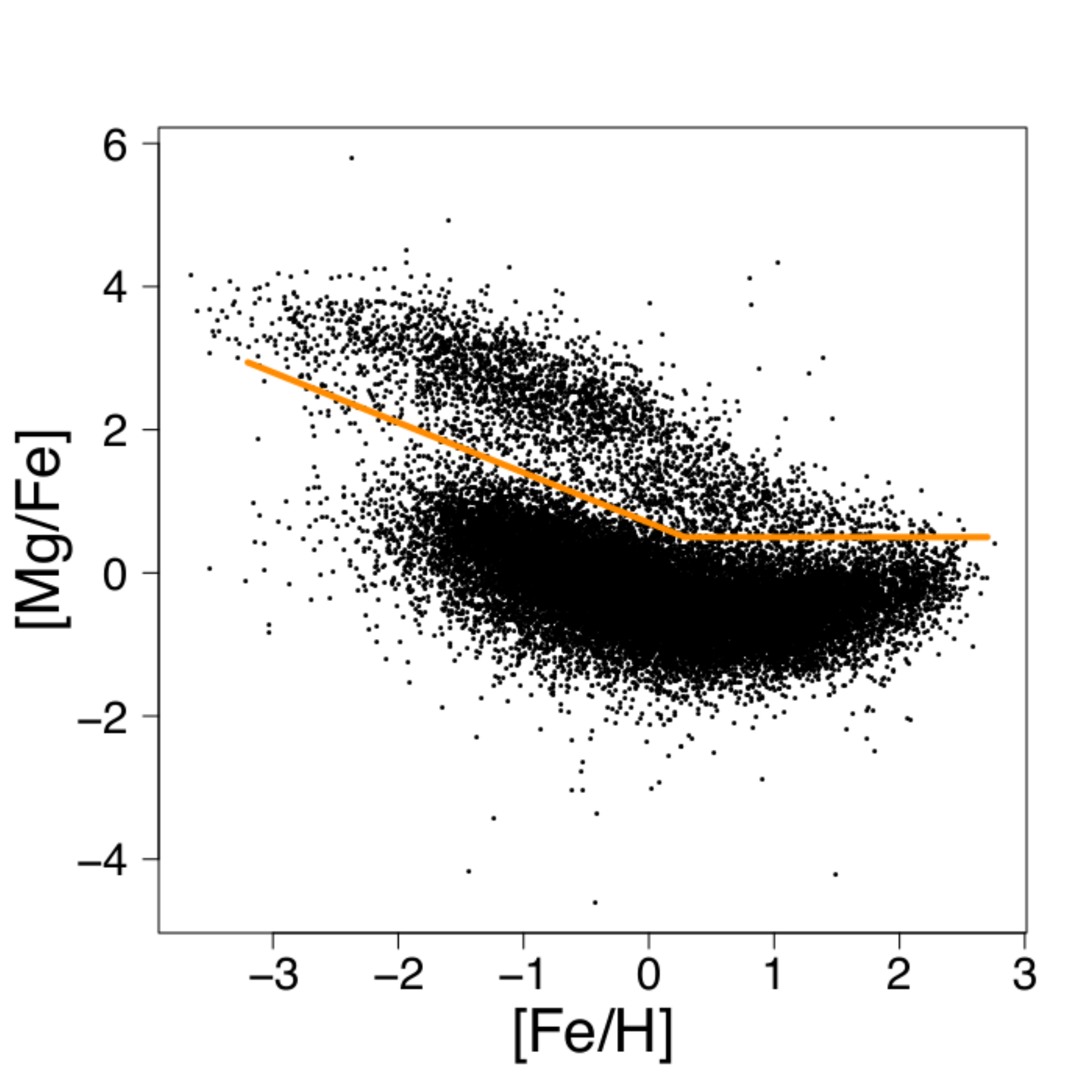}
\caption{The \num\ red clump stars in our sample projected into the standardized [Mg/Fe]-[Fe/H] plane. The expected bimodality associated with the high- and low-$\alpha$ sequences is evident. Typically, the two sequences are divided by eye. One such division of the stars is given by the orange line, with the high-$\alpha$ sequence being above and the low-$\alpha$ sequence being below. The traditional bend at \feh\ $\approx$ 0 used by \cite{Weinberg2018} and \cite{Blancato2019} outlines the low-$\alpha$ stars to create the recognizable banana-like shape. We will refer back to this reference split of the sequences many times throughout this paper.}
\label{fig:Alpha_Fe}
\end{figure}

Using Mg as our representative $\alpha$-element, the bimodality of our sample in the [Mg/Fe]-[Fe/H] plane is clearly visible in a scatter plot of the data presented in Figure \ref{fig:Alpha_Fe}. Following  \cite{Weinberg2018} and \cite{blancato2019variations}, we split the data by eye with two straight lines of differing slopes (orange line), joined  at [Fe/H] $\approx$ 0. Hereafter, we label the 3,211 stars above and the 23,924 stars below the orange line respectively as the high- and low-$\alpha$ sequence.

\begin{figure*}[]
  \centering
  \includegraphics[width=.95\linewidth]{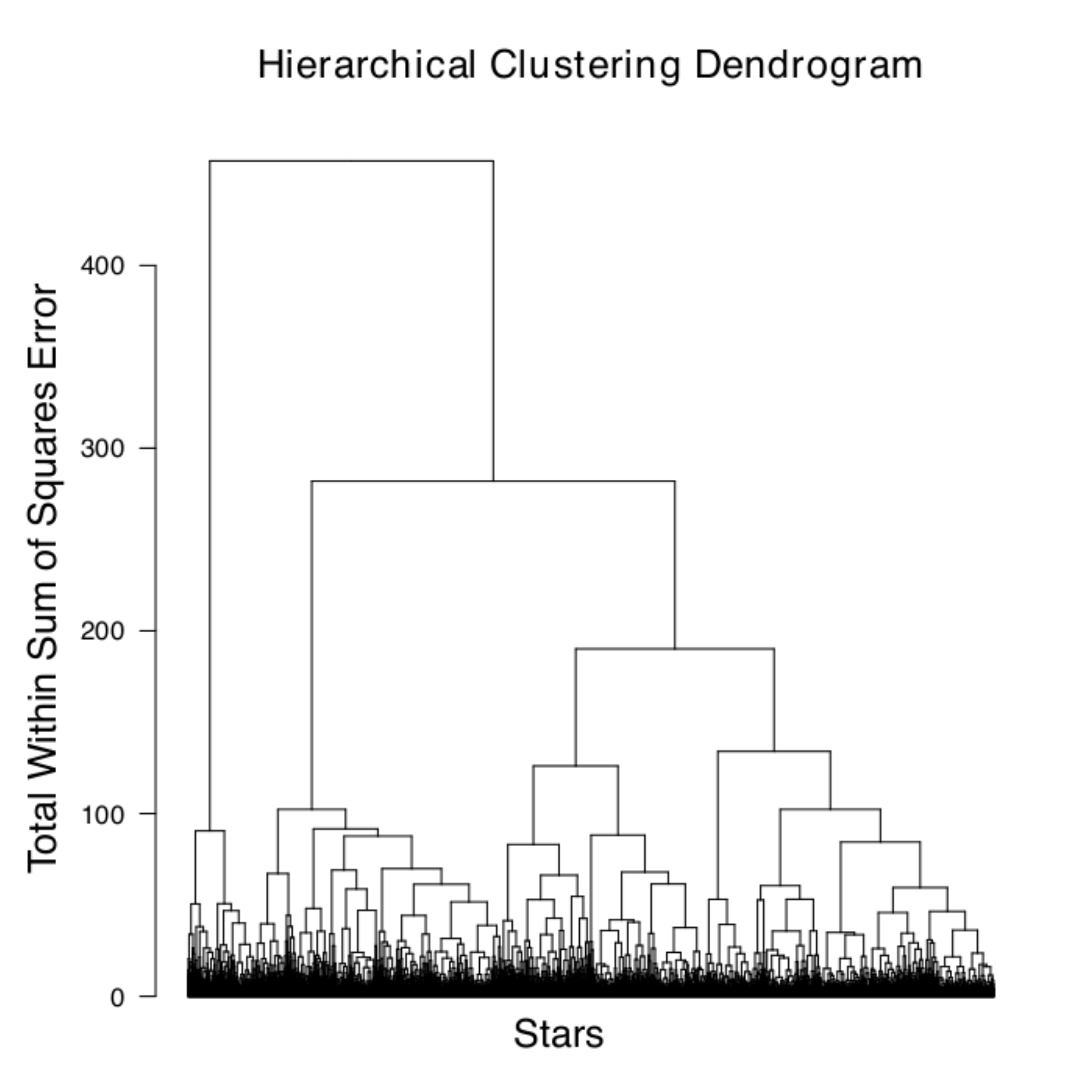}
  \caption{Dendrogram produced via agglomerative hierarchical clustering using Ward's Minimum Variance Criterion on \num\ RC stars. This dendrogram presents the clustering structure of the data, where the height of the y-axis represents the total within-cluster sum of squares (WSS) for the given number of clusters. The bottom (at height 0) is where each star is its own individual cluster and agglomerates to the top (at height $\sim$ 450) where where all the stars combine into one main cluster. In order to best determine the number of clusters our sample is comprised of, we compare the relative total WSS when successive groups combine. The total WSS for when there is one cluster is 457, and jumps down to 282 when the data split into two clusters. This jump is subjectively large, and justifies two main populations in our sample. However, the differences between total WSS for two through six clusters can all be considered ``large", and thus makes finding the number of meaningful populations in our dataset nontrivial.}
  \label{fig:dendrogram}
\end{figure*}

In this section we examine how two clusters defined in the full 19 dimensional chemical abundance space compare to the high- and low-$\alpha$ sequences determined by eye, after projection into the [Mg/Fe]-[Fe/H] plane. We then investigate the true dimensionality of the data, and cluster again in representative lower dimensional hyperplanes to demonstrate that the clusters from 19 dimensions are genuine. We additionally explore the sequences in many different abundance planes to determine the key elements contributing and informing the clustering. We differ exploration of age and spatial relationships until Sections \ref{sec:groups_age} and \ref{sec:groups_spatial}.

\subsection{Clustering the stars using their 19 abundance dimensions}
\label{sec:2Seq}
\label{sec:Hierarchical_results}

We first cluster our \num\ stars using their vector of 19 abundances with the hierarchical clustering method described in Section \ref{sec:ward}. We project the clustering hierarchy from individual stars to a single grouping in the dendrogram, shown in Figure \ref{fig:dendrogram}. The y-axis of the dendrogram represents the potential difference between clustered groups. As discussed in Section \ref{sec:ward}, the larger this difference, the more significant the group classification. To first examine the most simple clustering case, we find a large jump in the total within-cluster sum of squares (WSS) from 457.2 to 281.9 in going from one to two clusters, which splits the data into two samples of 2,292 and 24,843 stars, respectively.

\begin{figure*}
\centering
      \includegraphics[width=1.\textwidth]{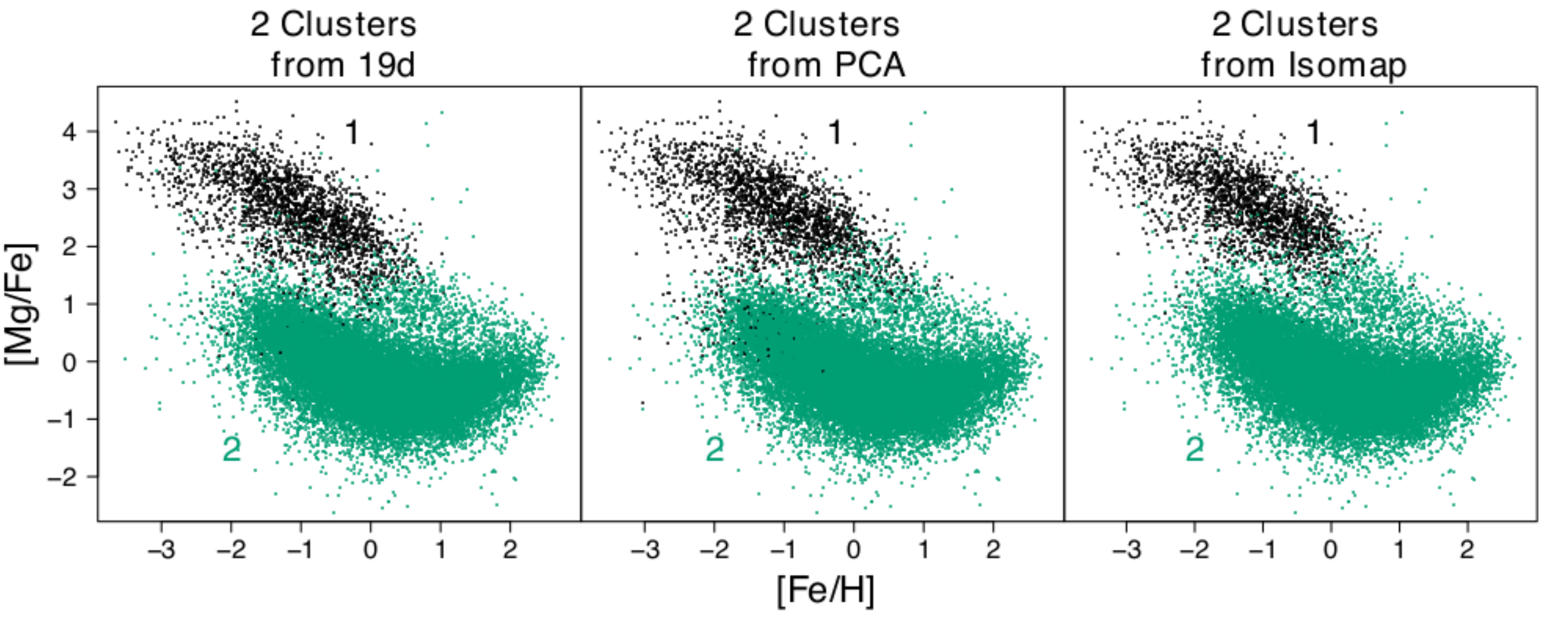}
\caption{The 2D projection in the standardized [Fe/H]-[Mg/Fe] plane of the two clusters that are found using agglomerative hierarchical clustering, with Ward's minimum variance criterion for the \num\ red clump stars, using \textbf{left:} the 19 dimensional chemical abundance space, \textbf{middle:} the first two dimensions of PCA decomposition, \textbf{right:} the first two dimensions of Isomap decomposition.  We refer to these two clusters as 1 (black) and 2 (green). In all three plots there is a group of stars between 0 $\leq$ [Fe/H] $\leq$ 2 and 0 $\leq$ [Mg/Fe] $\leq$ 1.5 that typically are visibly classified as high-$\alpha$ stars, but more closely resemble the low-$\alpha$ sequence in 19 dimensions and the first two main dimensions of lower dimensional embeddings. The stability of this grouping over the three different methods suggests that the results are a real yet unexpected feature of the data when using this algorithmic approach to clustering.}
\label{fig:Alpha_Fe_3}
\end{figure*}

The projection of these two clusters in the [Mg/Fe]-[Fe/H] plane is given as the leftmost plot in Figure \ref{fig:Alpha_Fe_3}. The two clusters look very similar to the previously defined high- and low-$\alpha$ sequences, with the black stars (cluster 1) corresponding to the high-$\alpha$ sequence and the green stars (cluster 2) corresponding to the low-$\alpha$ sequence and part of what is typically included as the high-$\alpha$ sequence (highlighted as yellow stars in top left of Figure \ref{fig:MisclassifiedStars}). While these two sequences are similar to the by-eye division, the subgroup of just over 900 stars with $0 \leq$ [Fe/H] $\leq 2$ and $0.5 \leq$ [Mg/Fe] $\leq 1.5$ is assigned to the same cluster as the bulk of the low-$\alpha$ stars. 

\subsection{Correlation structure of the data}
\label{sec:our_data}

\begin{figure*}[]
  \centering
  \includegraphics[width=0.8\linewidth]{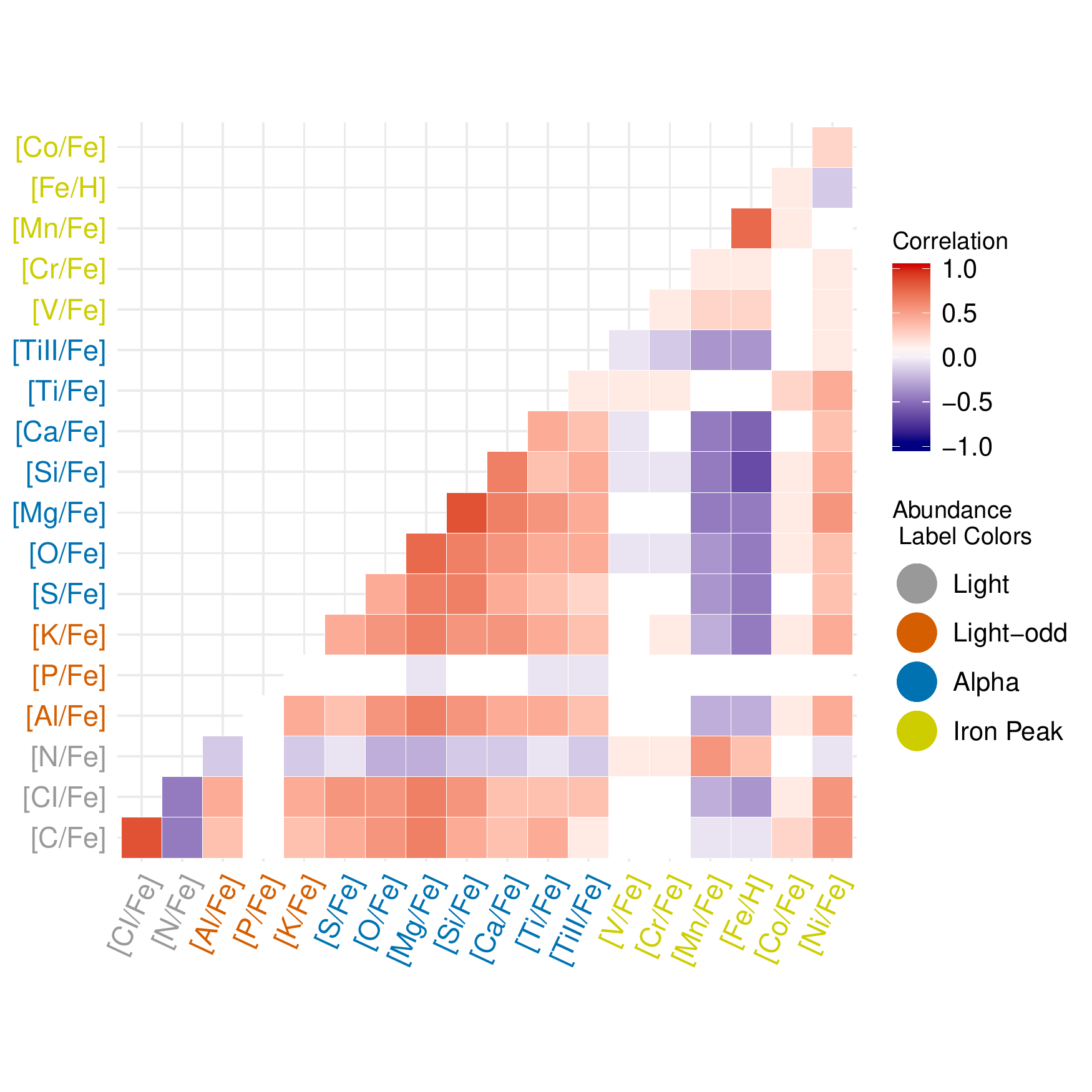}
  \vspace{-25pt}
  \caption{The correlation structure in the standardized abundances from the \apogee\ red clump catalogue for \num\ stars. The correlations are quantified using the scale shown at right. The abundance labels are coloured according to their nucleosynthetic family and the elements are ordered within these families. Primarily, families either have positive correlations among their members (i.e. $\alpha$- and iron-peak elements), or they have strong correlations that are both positive and negative (i.e. light elements). The strong intra-family relationships are indicative that the true dimensionality of the data is $<$ 19 dimensions of the individual abundances that are measured, and motivates our use of dimension reduction techniques.}
  \label{fig:cors}
\end{figure*}

In the previous subsection, we showed the two main clusters in 19 dimensions projected into the two dimensional [Fe/H]-[Mg/Fe] plane. In order to determine the validity of the groupings found in 19 dimensions, we continue our analysis by investigating the structure of our data sample. Examining the correlation structure of each pair of abundances allows us to see if the data lie in a higher dimensional ambient space than needed for doing work on the data such as clustering. Since our variables are normalized to have 0 mean and a standard deviation of 1, the correlation coefficient between abundance vector $x$ and abundance vector $y$ is $$r_{x,y} = x \cdot y$$ where $n = \num$.

Figure \ref{fig:cors} shows the correlation structure of the data that we calculate, in a matrix organized by elements divided into their families. A darker shade of either red or blue illustrates a stronger positive or negative linear relationship between the corresponding abundances, whereas a light shade indicates a very weak linear relationship between the two. In order to determine if there are any intra-family relationships, we categorize the abundances into four groups according to how the elements were produced: light elements with even atomic number (C, CI, N), light elements with odd atomic number (Al, P, K), $\alpha$-elements (S, O, Mg, Si, Ca, Ti, TiII), and iron-peak elements (V, Cr, Mn, Fe, Ni). Figure \ref{fig:cors} reveals that there does appear to be trends within the different familial groups. The $\alpha$-elements are fairly strongly positively correlated with one another, while the iron-peak elements are only weakly correlated with each other. On the other hand, the three light elements with even atomic numbers are very strongly linearly related. Looking at specific elements, [N/Fe], [Fe/H], and [Mn/Fe] are anti-correlated with nearly every other abundance. It is also apparent that [P/Fe] has no strong correlation with any of the abundances. Similarly, [V/Fe] and [Cr/Fe] are not strongly correlated with other elements, however they are somewhat linearly related with other iron-peak elements.   

Due to some elements having quite strong linear correlations, we see from this simple investigation that the true dimensionality of the data is lower than the number of elements measured and considered in this work. Therefore, we proceed with implementing dimensionality reduction techniques in our clustering approaches.

\subsection{Investigating Lower Dimensional Embeddings}
\label{sec:lowerDim}

\subsubsection{Exploring the Influential Dimensions}
\label{sec:pca_components}

\begin{figure*}[]
  \centering
  \includegraphics[width=0.95\linewidth]{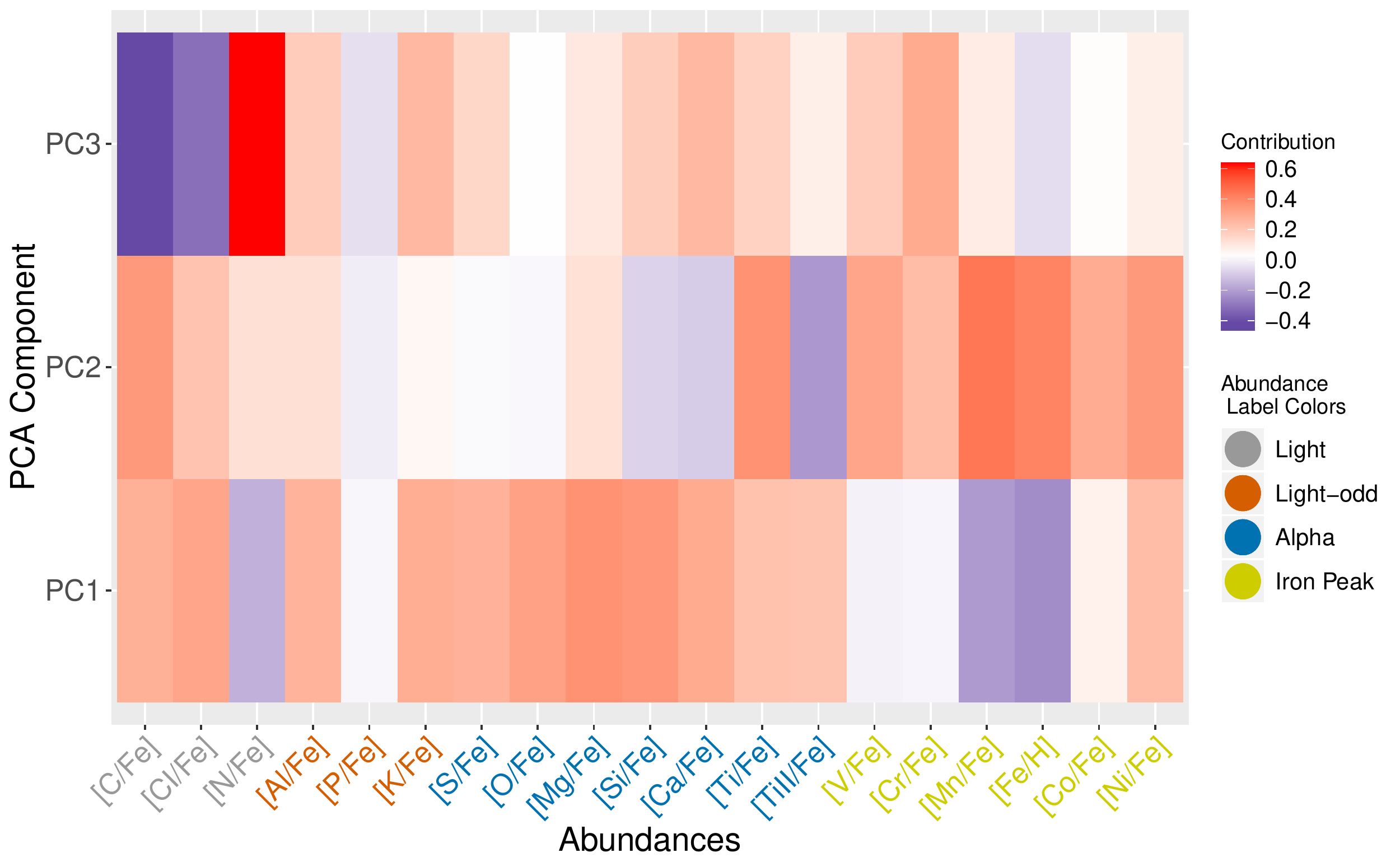}
  \caption{Depiction of how much each of the 19 chemical abundances contributes to the first three principal components. The first three components explain just over 50\% of the variance of the \num\ RC stars in the 19 dimensions. Grouping and coloring the abundances by nucleosynthetic family allows us to quickly identify that the $\alpha$-elements all positively contribute to the first component, the second component primarily captures the iron-peak elements, while the light elements with even atomic number contribute to the third.}
  \label{fig:PCA_contributions}
\end{figure*}

Section \ref{sec:our_data} demonstrated that the abundances are highly correlated. Therefore, we now implement PCA on our 19 dimensional abundance data for our 27,135 stars to identify the primary features that dominate the clustering hierarchy. We find that the first three principle components explain 53\% of the variance in the data. Therefore, focusing on these components allows us to examine how the abundances contribute to the key aspects of the data's distribution in higher dimensions. The relative contributions of each element to the first three principal components is shown in Figure \ref{fig:PCA_contributions}. This figure demonstrates that the $\alpha$-elements all positively contribute to the first principal component (PC1), the iron-peak elements all positively contribute to the second principal component (PC2), and the light elements with even atomic number are the main contributors to the third principal component (PC3). [P/Fe] does not strongly contribute to any of the first three components, rendering it insignificant in analyzing the dataset's variability. It should also be noted that [Mg/Fe] provides the most absolute contribution to PC1, which is reassuring of [Mg/Fe] being a sound representative or proxy for an overall $\alpha$-element in considering the $[\alpha$/Fe]-[Fe/H] plane.

\begin{figure*}[]
  \centering
  \includegraphics[width=.45\linewidth]{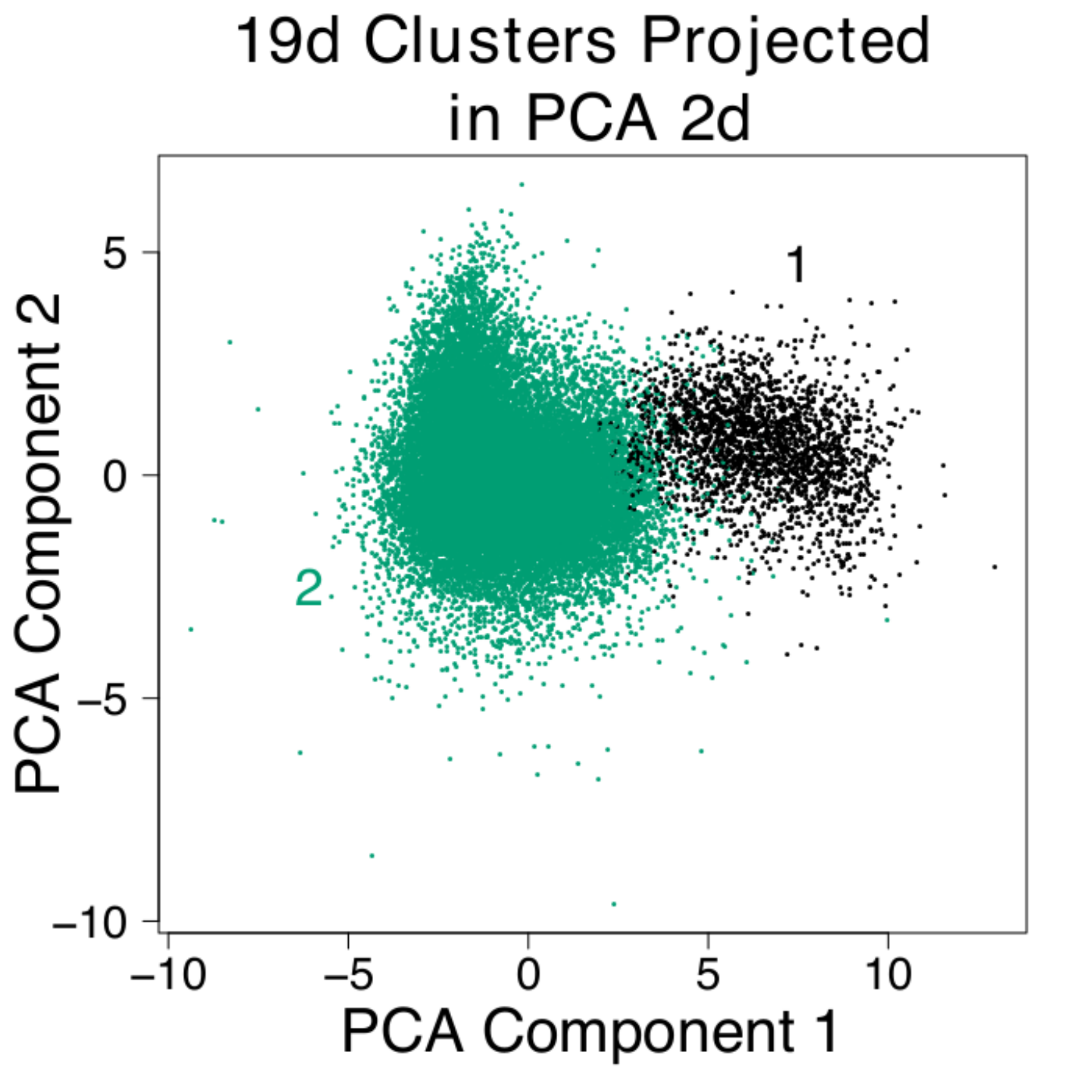}
  \includegraphics[width=.45\linewidth]{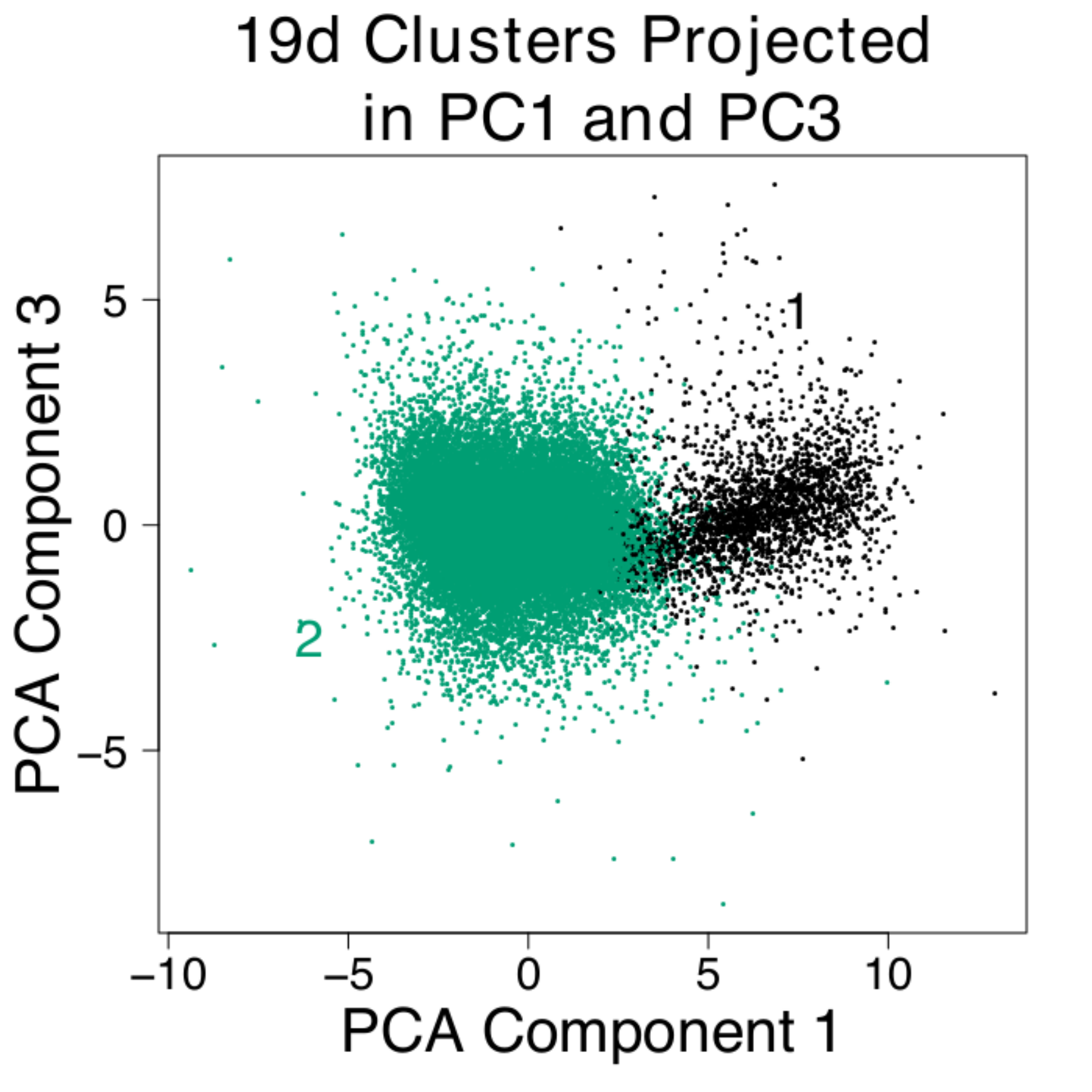}
  \caption{\num\ RC stars projected into \textbf{left:} the first two components of PCA, which explain 46\% of the variance in the 19 dimensional data, \textbf{right:} the first and third PCA components. The stars are color coded according to their clusters found in 19 dimensions using agglomerative hierarchical clustering with Ward's minimum variance criterion. These clusters coincide with the two noticeable groups created in both planes. Since the first three components capture more than half of the data's variability from 19 dimensions, our choice of first examining two populations is validated.}
  \label{fig:PCA_Plane}
\end{figure*}

We now examine the three principal components from Figure \ref{fig:PCA_contributions} in two-dimensional projections. We show the first two components and the first and third components in the left and right of Figure \ref{fig:PCA_Plane} respectively to validate the significance of examining two groups. In both projections, the stars (colored according to their grouping defined in Section \ref{sec:Hierarchical_results}) are split into two primary populations, with each group corresponding to either the black (cluster 1) or green (cluster 2) cluster with little cross over. In addition to being spatially discrete in these planes, the two sequences show different behaviors in these projections. The black cluster (cluster 1) is evenly distributed along PC1 and PC2 and shows a strong positive correlation between PC1 and PC3, while the green cluster (cluster 2) is evenly distributed amongst all three components. This confirms the populations contain distinct differences. 

\subsubsection{Determining the Validity of the Clustering Model}

One concern in working with high-dimensional data is the risk of over-fitting due to the curse of dimensionality. In order to determine if the group of $\sim 900$ stars discussed in Section \ref{sec:Hierarchical_results} are ``misclassified" as a side effect of working in 19 dimensions, we cluster the stars using only the first two principal components (PC1, PC2) rather than their 19 abundances as previously done. We choose to only use the first two components as PC1 and PC2 together explain 46\% of the variance of the data, thus capturing nearly half of the data's spread and ensuring we only examine the main features of the data. We again project the resulting two clusters back into [Mg/Fe]-[Fe/H] plane as shown in the middle panel of Figure \ref{fig:Alpha_Fe_3}. The results are almost identical to clustering in 19 dimensions --- 89\% of the small group of $\sim$ 900 stars in the green cluster (cluster 2) that join the black cluster (cluster 1) at high-$\alpha$ values are similarly identified as being part of the green cluster (cluster 2) using the first 2 principal components. This shows that this group of stars differentiates from the high-$\alpha$ sequence along the linear axes that describe the most spread.

\begin{figure}[]
  \centering
  \includegraphics[width=.95\linewidth]{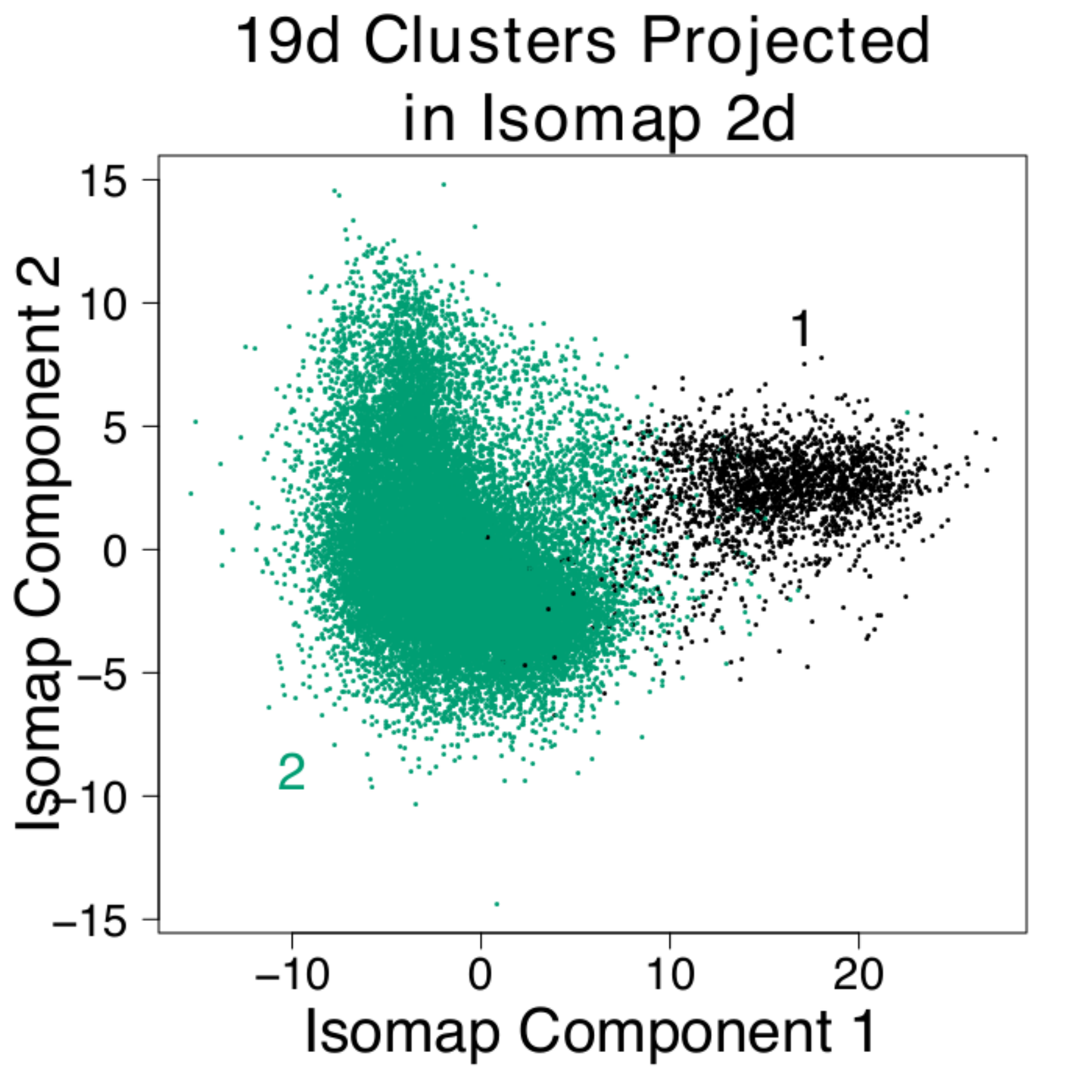}
  \caption{The RC \apogee\ sample of \num\ stars projected into the first two components of Isomap. The stars are colored according to their sequences from 19 dimensions, determined via agglomerative hierarchical clustering with Ward's minimum variance criterion. Similar to PCA, the first two components split primarily into two groups, further confirming that two groups is reasonable for this dataset.}
  \label{fig:Isomap_Plane}
\end{figure}

In 19 dimensions we are unsure of the structure of the data, and a linear projection (eg. PCA) may not be optimal. Therefore, we also run a non-linear dimension reduction technique to compare our results with this methodology. This is the Isomap reduction discussed in Section \ref{sec:Isomap}. It is clear from  Figure \ref{fig:Isomap_Plane} that the first two components of Isomap look analogous to the first two components of PCA, demonstrating the merit of clustering into two primary groups that is consistent between different methodologies and assumed models.  

Following the same approach as with PCA, we choose to cluster the stars in the first two dimensions of Isomap to examine if the two clusters are differently distributed and what happens to the group of metal-rich cluster 2 stars from the left of Figure \ref{fig:Alpha_Fe_3}, that are often associated with the high-$\alpha$ sequence (cluster 1). The two clusters determined with the data following Isomap dimensionality reduction are projected back into the [Mg/Fe]-[Fe/H] plane in the far right panel in Figure \ref{fig:Alpha_Fe_3}. Yet again, the two clusters show the same distribution as with PCA and without dimensionality reduction. The results are again almost identical to clustering in 19 dimensions --- 97\% of the small group of $\sim$ 900 stars in the green cluster (cluster 2) that join the black cluster (cluster 1) at high-$\alpha$ values are similarly identified as being part of the green cluster (cluster 2) using the first 2 first Isomap dimensions.

\subsection{Projecting our two clusters across different abundance planes}

We want to examine the projection of the two largest clusters from hierarchical clustering in different abundance planes. In particular, we are interested in the subset of metal-rich stars that have been typically associated with the high-$\alpha$ sequence (see Figure \ref{fig:Alpha_Fe}) yet we find associated with our green cluster (cluster 2; Figure \ref{fig:Alpha_Fe_3}), which is comprised of low-$\alpha$ stars. We therefore differentiate this group of stars by eye from the rest of the green cluster (cluster 2) in a series of two-dimensional abundance planes to demonstrate that they sensibly align with the group the algorithm associates them with.

\begin{figure*}[]
  \centering
  \includegraphics[width=.35\linewidth]{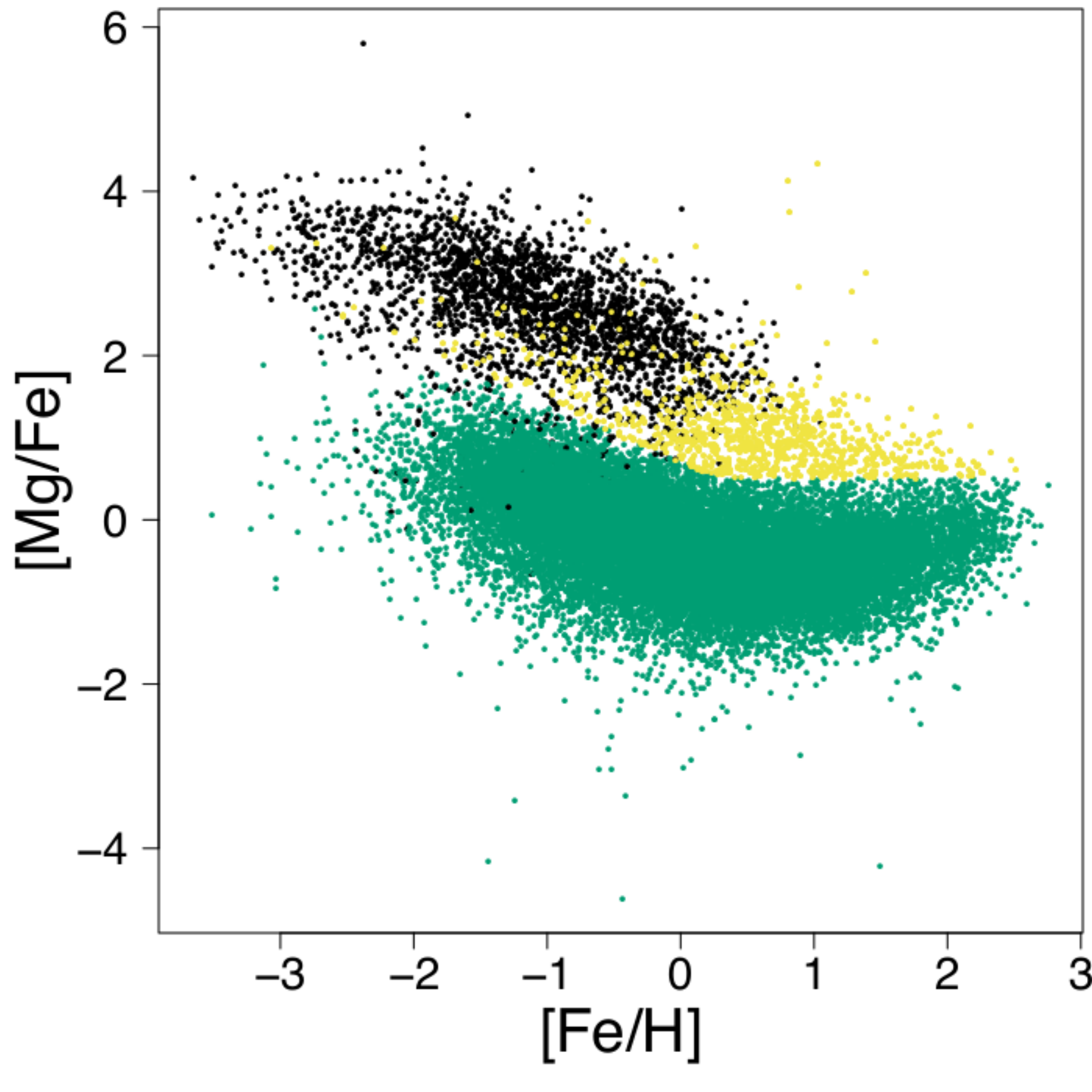}
  \includegraphics[width=.35\textwidth]{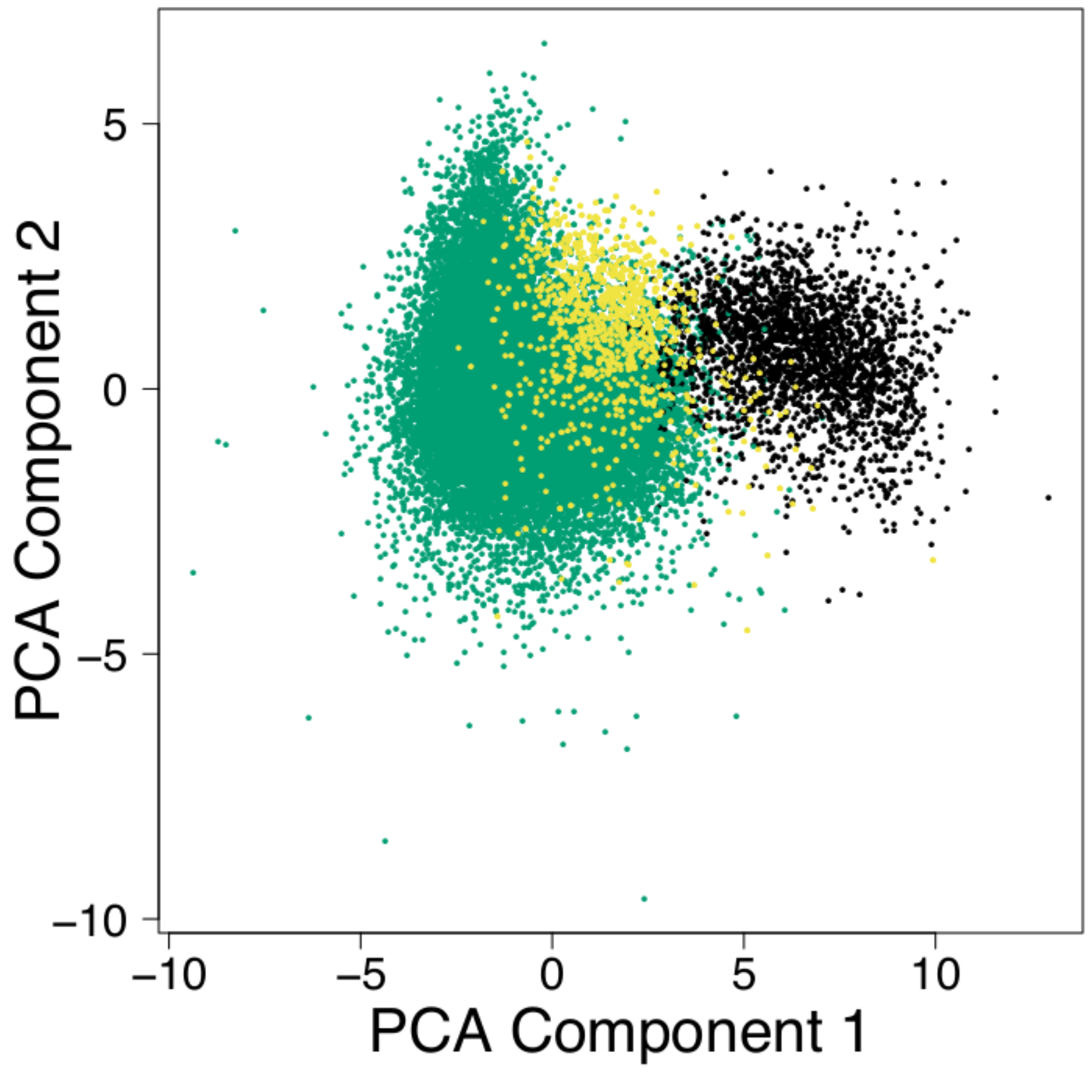}
  \includegraphics[width=.35\textwidth]{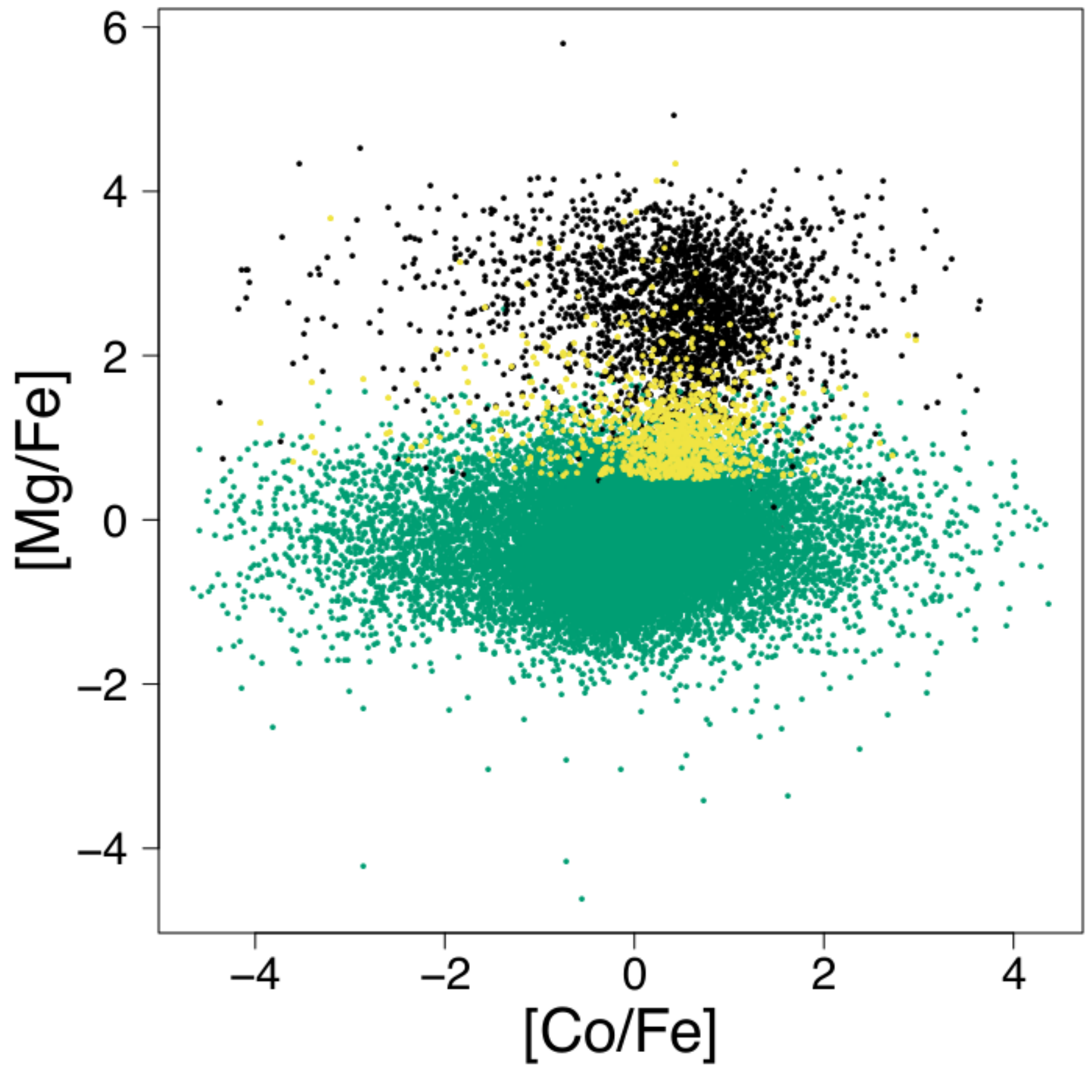}
  \includegraphics[width=.35\textwidth]{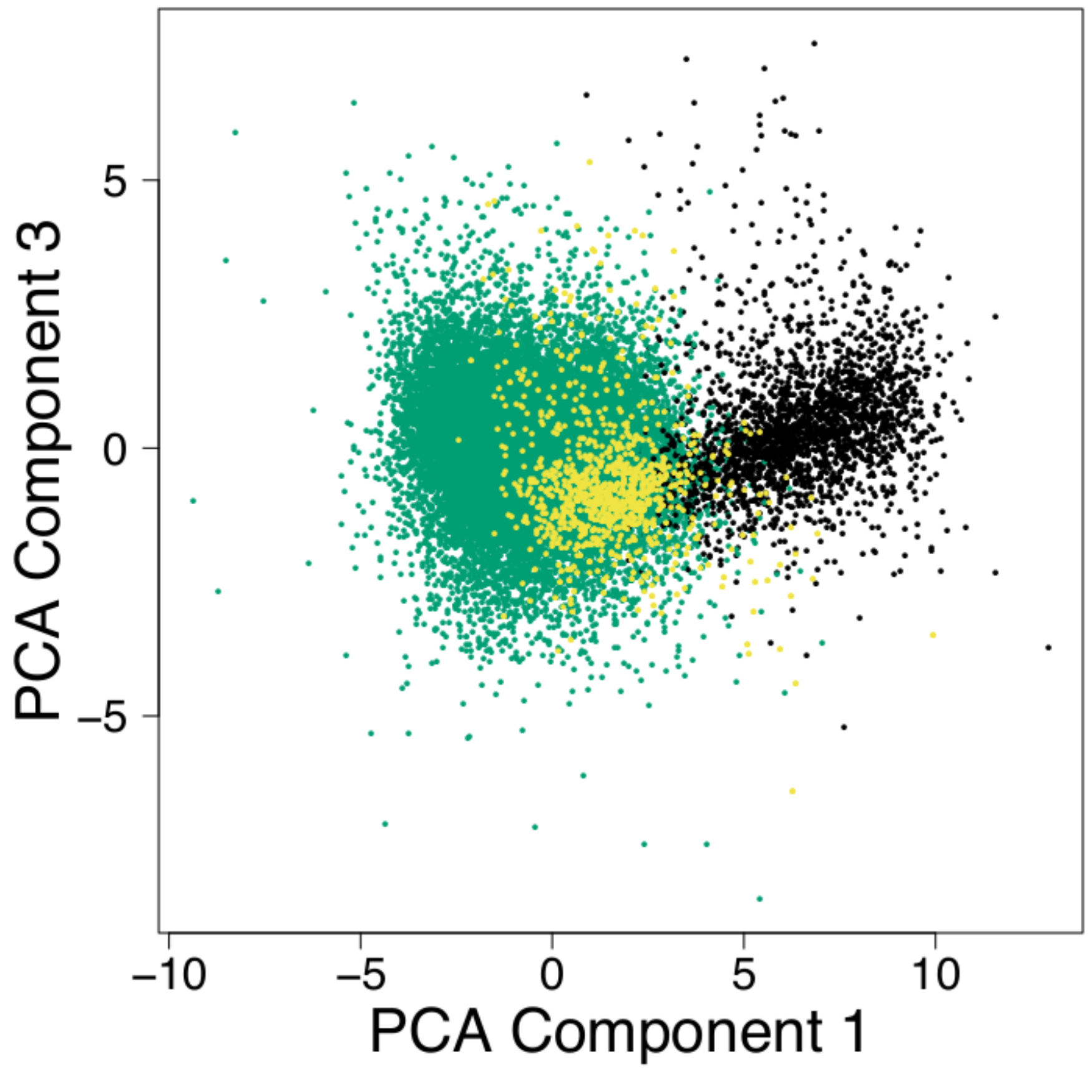}
  \includegraphics[width=.35\textwidth]{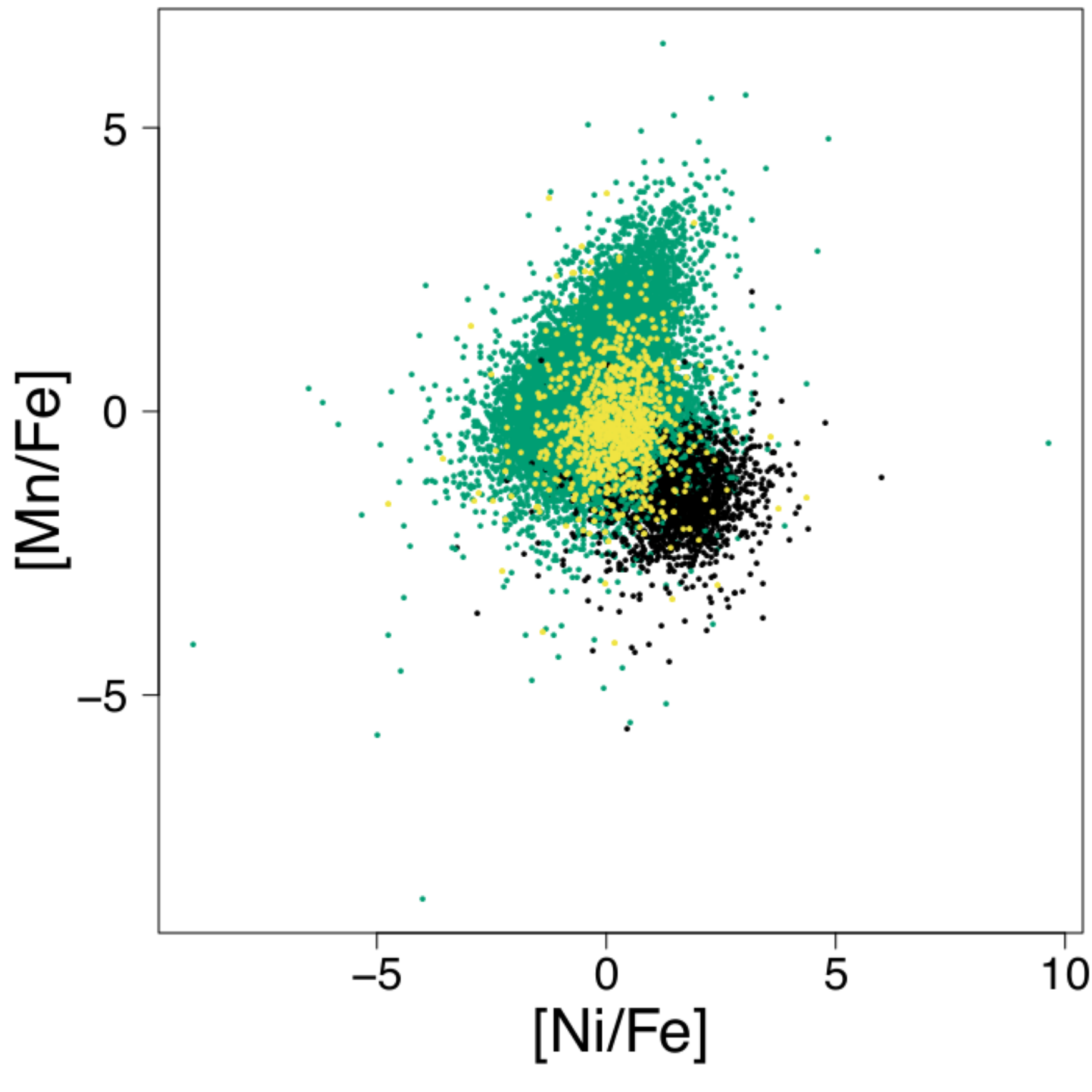}
  \includegraphics[width=.35\textwidth]{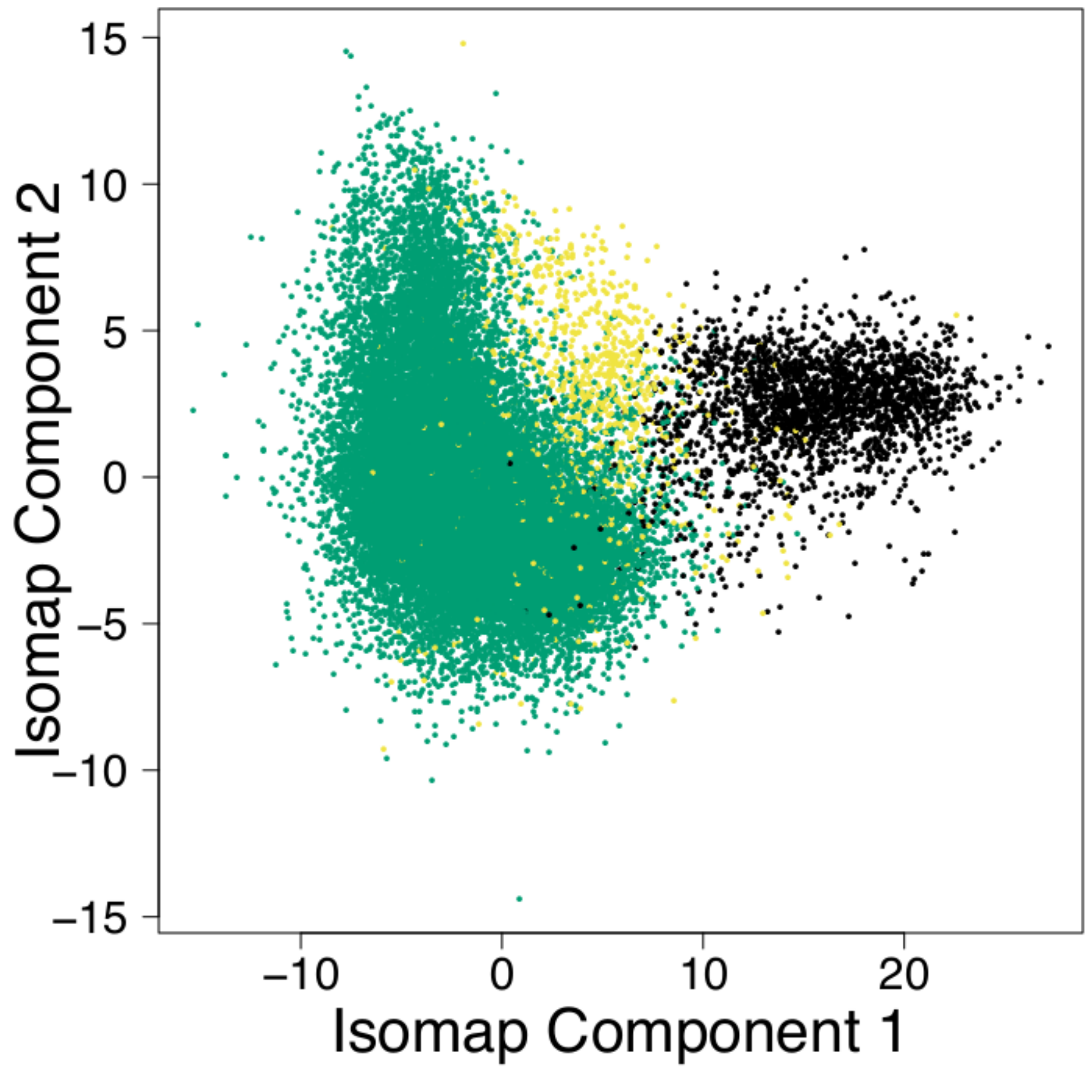}
  \caption{\num\ RC stars projected into \textbf{left}: three abundance-abundance planes, \textbf{right}: the first three components of PCA and first two dimensions of Isomap, in order to investigate where the $\sim$ 900 low-$\alpha$ stars (colored in yellow) live in different planes. The black group is the (higher-$\alpha$) sequence of stars (cluster 1) from the left of Figure \ref{fig:Alpha_Fe_3}, while the green and yellow groups combine to create the green cluster (cluster 2) from the same figure. These plots were chosen as representatives from the full set of abundance-abundance planes. In some, there are visibly two groups of stars, where the yellow stars are part of the green group where the black group connects (middle left). In others, the yellow stars are dispersed throughout the green group (bottom left). The first two dimensions of PCA and Isomap reveal that this group of stars are unique --- closer resembling the green group but they are in a distinct, less dense region than the rest of the green stars. Similar to some abundance-abundance planes, PC1-PC3 shows the yellow group as an extension of the black group into the green group.}
  \label{fig:MisclassifiedStars}
\end{figure*}

Specifically, we are interested in the yellow stars in Figure \ref{fig:MisclassifiedStars}. After going through the ${19 \choose 2}$ abundance-abundance plots, Figure \ref{fig:MisclassifiedStars} contains the unique plots that show the different scenarios in which the group of stars that are typically classified as high-$\alpha$ stars (black cluster) more closely resemble the low-$\alpha$ sequence (green cluster). Additional abundance-abundance plots are given in the Appendix. In many of the plots, there are two blob-like structures, such as in the middle left plot. Here one group contains the black stars while the other mainly contains the yellow and green stars. Visually, the yellow stars are considered to be part of the green population, but they are located where the black group connects to the green one. In the bottom left plot we see that the two groups are less distinct, but here the yellow stars are condensed, located in the center of the green group. We also show at the top and bottom right of Figure \ref{fig:MisclassifiedStars} where the group of $\sim$ 900 stars live in the first two dimensions of PCA and Isomap. In both cases, but more strikingly in the Isomap plane, the stars are located in the less dense region of the green cluster near the connection to the black cluster. This suggests that while these stars do more closely resemble the low-$\alpha$ sequence, they also contain qualities which make them unique to both clusters. From the PCA1-PCA3 plane (center right) the yellow stars appear to be an extension of the black cluster, but again located within the green group.

\begin{figure*}[]
  \centering
  \includegraphics[width=.95\linewidth]{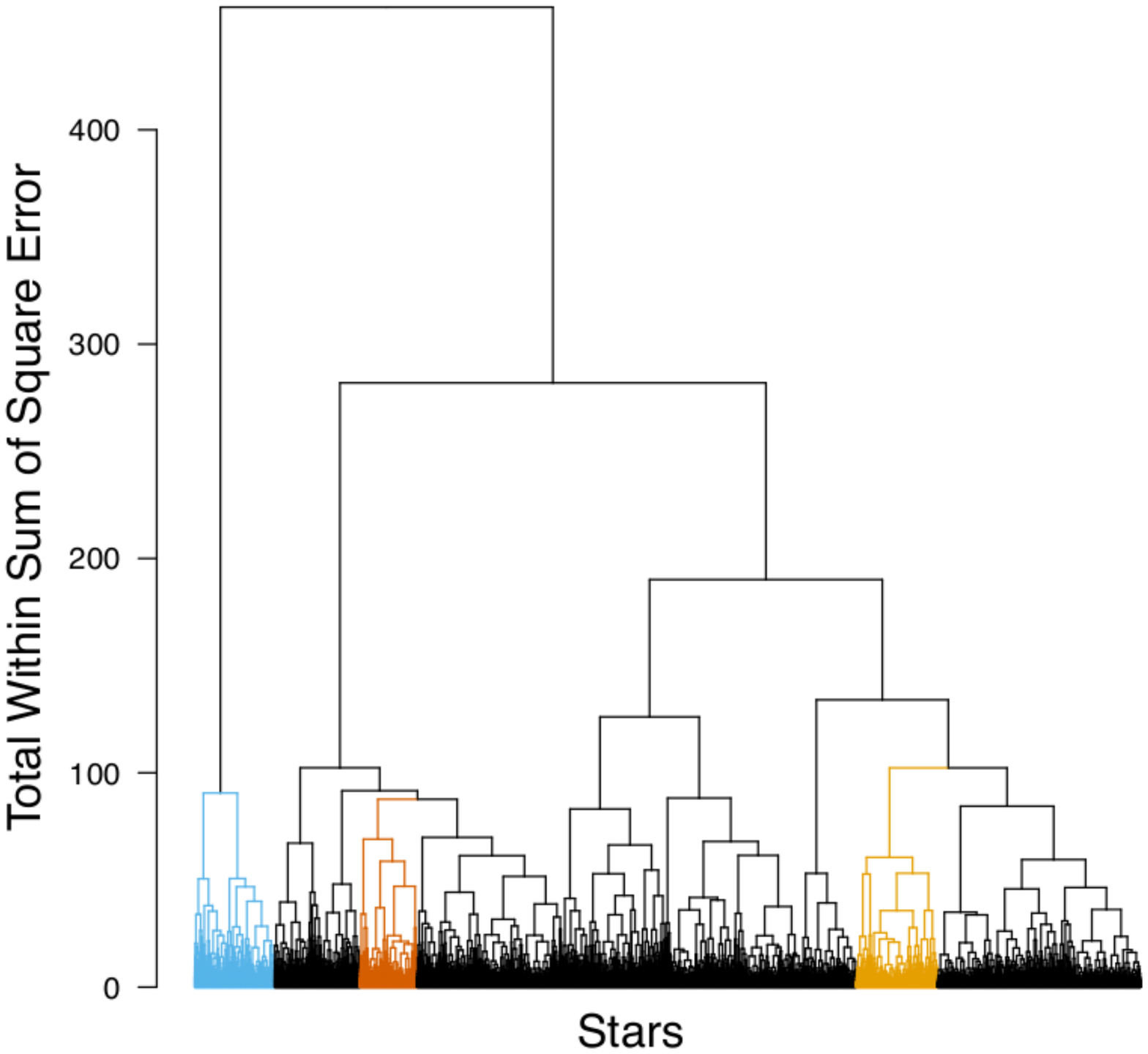}
  \vspace{-30pt}
  \caption{19 dimensional clustering dendrogram for \num\ RC stars created using agglomerative hierarchical clustering with Ward's minimum variance criterion. The two branches marked contain the majority of the $\sim$ 900 low-$\alpha$ stars that typically are classified as part of the high-$\alpha$ sequence in the [$\alpha$/Fe]-[Fe/H] plane. The red-orange group contains $\sim$ 42\% of this group of stars, followed by the orange group containing $\sim$ 20\%. The other 38\% are spread evenly throughout the other branches of the larger cluster. This shows that these stars contain large differences among each other, and do not show any resemblance in 19 dimensions to the more metal-poor high-$\alpha$ stars (cluster 1 in left of Figure \ref{fig:Alpha_Fe_3}), which are highlighted here as light blue.}
  \label{fig:Misclass_dend}
\end{figure*}

The natural question that follows on from this is if this small subset of metal-rich stars is similar throughout the hierarchy of clusters, or contains distinct branches within itself. Figure \ref{fig:Misclass_dend} shows where the majority of the green clustered (cluster 2) metal-rich stars lie in the dendrogram. This subset of stars splits nearly in half when the entire sample of stars divides into three clusters, revealing that there are major differences within this unique group of $\sim$ 900 stars. Additionally, the majority of these stars lie within clusters with fairly small total WSS near 100, which is a very large jump to where the high-$\alpha$ sequence combines at a total WSS of 457. Between this and the fact that these stars are not all grouped together shows that these stars are well separated from the high-$\alpha$ sequence in the 19 dimensional chemical abundance space. This combined with the pairwise abundance plots indicates that a division of the abundance distribution into two clusters only is perhaps sub-optimal. That is, the data can be better described as breaking up into more than two groups.

\section{Results II: number and properties of significant clusters}
\label{sec:groups}

We have established that the data justifiably clusters into at least two groups using different approaches. Additionally, we have shown that these two groups are different than those typically visually assigned in the [Fe/H]-[$\alpha$/Fe] plane.  The next question we ask is how many underlying clusters are our data potentially comprised of, and how do we determine this number? Given the dendrogram in Figure \ref{fig:dendrogram}, two, three, four, and six clusters all seem reasonable for this dataset given the large difference in the total WSS measurement on the y-axis. The second row of Figure \ref{fig:All_Clusters} shows the split in the dendrogram for potential clusters for two through six groups. The top row shows the respective breakdown in the [Mg/Fe]-[Fe/H] plane.

\begin{figure}[]
  \centering
  \includegraphics[width=.95\linewidth]{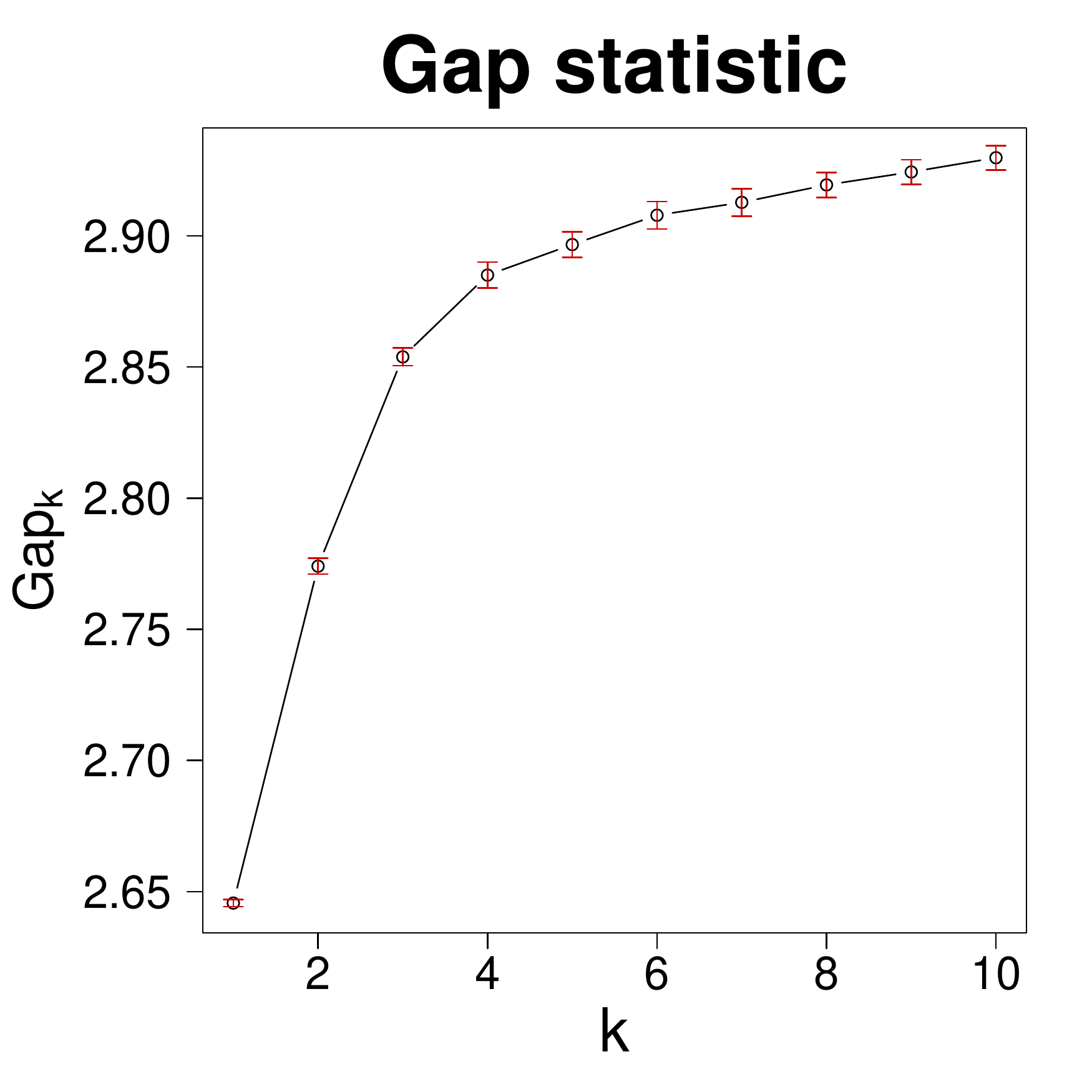}
  \caption{The Gap statistic curve created using 10 bootstrapped samples of size \num\ with 1-$\sigma$ error bars. The Gap statistic helps determine the number of clusters based on the 19 dimensional abundance space alone. The Gap statistic for six clusters only just meets the nominal criteria for deciding the highest number of clusters the data can be decomposed into, of being larger than the previous cluster's lower error bar. This is primarily due to the small error bars from our large sample size. This statistic, however, does suggest that there are at least six underlying populations in the data, but the Gap statistic for four through ten clusters are all very close.}
  \label{fig:GapStat}
\end{figure}

As mentioned previously in Section \ref{sec:Hierarchical_results}, we compare the total WSS to infer the number of groups. Splitting the data into one, two, three, four, five and six clusters yields a total WSS of 457.2, 281.9, 190.2, 134.1, 126.1, and 102.3 respectively. To help determine the most appropriate cutoff, we examine the Gap statistic introduced in Section \ref{sec:gapDef}. The associated plot given in Figure \ref{fig:GapStat} shows the Gap statistic, with 1-$\sigma$ standard deviation error bars for one through ten clusters. The error bars are small primarily due to the large sample size. This is why we choose to run only 10 bootstrapped samples; increasing the number of bootstraps makes the error bars even more insignificant. The Gap statistic for four through ten clusters are all very close to reaching the significance threshold (a flattening of the statistic) but the turn-over is around six clusters. This suggests that the data can be meaningfully represented by up to six clusters. However, more generally, the roll of this statistic is indicative of a continuum of structure underlying the abundance distribution.

\subsection{Ages as tags of cluster populations}
\label{sec:groups_age}

\begin{figure*}[]
  \centering
  \includegraphics[width=.95\linewidth]{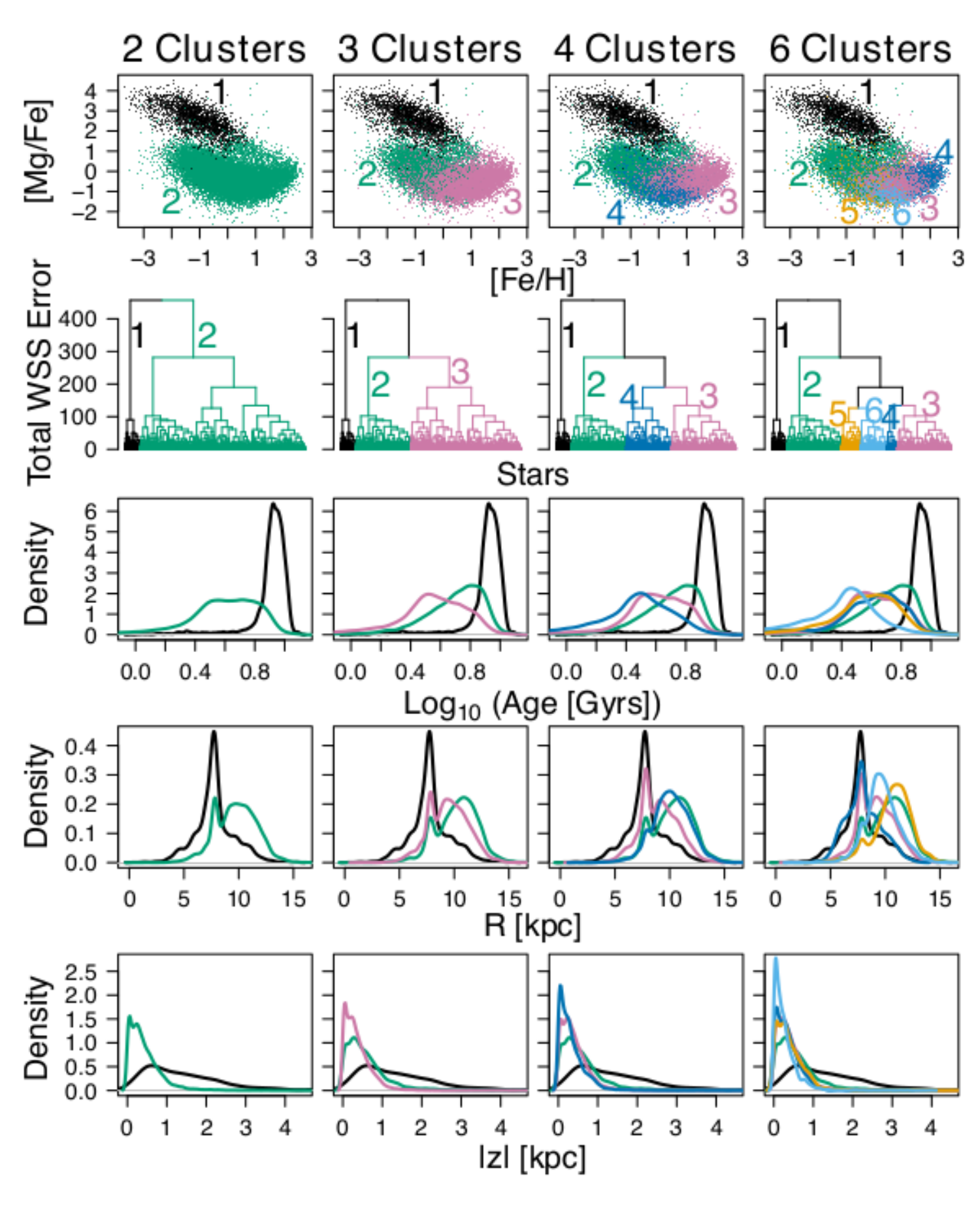}
  \caption{The colored \textbf{top row:} [Mg/Fe]-[Fe/H] planes, \textbf{second row:} dendrograms, \textbf{third row:} log$_{10}$ age densities, \textbf{fourth row:} \rgal\ densities, \textbf{bottom row:} $|z|$ densities for two, three, four, and six clusters. The colors order the clusters by mean age, with black (cluster 1) labeling the oldest group, then green (cluster 2), red (cluster 3), blue (cluster 4), orange (cluster 5), and finally light blue (cluster 6) defining to the youngest group of stars.}
  \label{fig:All_Clusters}
\end{figure*}

We now wish to test if the different clustered groups are also groups with different mean times of formation. Presumably, if the clustering of abundances links in any physically meaningful way to how the disk was formed over time, we might expect our statistically identified populations to have discrete ages. We find that our clusters do have different mean ages, with age distributions for the two to six clusters shown in the third row of Figure \ref{fig:All_Clusters}. Therefore, we now test the age separation as an independent metric via which we can justifiably decompose the data into discrete clusters. 

The third row of Figure \ref{fig:All_Clusters} presents the $\log_{10}$ age densities for each cluster as a function of the number of total groups. The black population of stars (cluster 1) remains unaffected when separating the data up to six clusters. It is comprised of primarily the oldest stars from the sample, and its age distribution is noticeably different from the other clusters. For the data to divide into three groups, the green group from two clusters (cluster 2; top row) splits into a left and a right skewed distribution which peak at the edges of the original plateau. Then the youngest stars break away from the right skewed red cluster (cluster 3) to create four total clusters. Once the current blue group (cluster 4) separates, it becomes difficult to differentiate the middle aged clusters. For instance the black (cluster 1), green (cluster 2), and light blue (cluster 6) densities are easy to distinguish for six populations, but the orange (cluster 5), blue (cluster 4), and red (cluster 3) densities are quite similar. Despite having similar densities, the pairwise multiple comparisons test outlined in Section \ref{sec:ttest} performed on the $\log_{10}$ ages from \cite{Sanders2018} suggest that our data separate into six populations at a confidence level of 0.95. 

To investigate the sampling confidence of this result, we run 1,000 Monte Carlo simulations by drawing new ages for each star, $i$, from a Normal distribution with mean $\log_{10}$ age$_i$ and its associated error as the 1-$\sigma$ standard deviation. We report that 24.1\%, 31.4\%, and 44.5\% of the simulations respectively produced four, five, and six clusters for a confidence level of 0.95. This implies that for up to six populations the stars separate into groups of different mean ages. This does not necessarily imply that there are six underlying clusters physically in the data, but rather age is a fundamental single variable by which the stars are linked together via their many abundances. Despite the age distributions being difficult to differentiate in Figure \ref{fig:All_Clusters}, our sample, clustered with abundances only, suggests that each of the six clusters has a statistically significant different mean age. This leads us to next examine the spatial distributions of the clusters.  

\subsection{Spatial distributions of the  cluster populations}
\label{sec:groups_spatial}

We now examine the spatial distributions of our clustered groups by looking at the Galactocentric radial distance, \rgal, and the absolute distance from the mid-plane, $|z|$, (both in kpc and calculated assuming the Sun is 8 kpc from the Galactic center and 25 pc above the mid-plane). The bottom two rows of Figure \ref{fig:All_Clusters} show that for two to six clusters, the clusters have different distributions in \rgal\ and $|z|$.  The structure in these distributions is driven by the \apogee\ survey selection function (for example the relatively large number of stars in the specially targeted Kepler field near the Sun). It is the relative dissimilarity between the groups that is relevant and noteworthy here. This is indicative that stars that are most chemically similar (according to our clustering) are also more spatially similar to one another and different between groups.

Starting with two clusters (left of Figure \ref{fig:All_Clusters}), we see the black cluster (cluster 1) is strongly peaked in \rgal\ (fourth row) at \rgal\ $\approx$ 8 kpc while the green cluster (cluster 2) is bimodal with a sharp peak at \rgal\ $\approx$ 8 and broader peak from \rgal\ $\approx$ 9-11 kpc. To create three clusters, the green cluster (cluster 2) divides into two bimodal distributions, with the new groups having their second peaks at \rgal\ $\approx$ 9 (red cluster, cluster 3) and 11 kpc (green cluster, cluster 2). The red group (cluster 3) then splits to create four clusters, with the new red group (cluster 3) taking on the majority of the \rgal\ $\approx$ 8 kpc stars compared to the blue group (cluster 4). Six groups become difficult to discern, with the majority of the clusters losing their bimodality and becoming somewhat unimodal. 

The $|z|$ distributions (bottom row of Figure \ref{fig:All_Clusters}) allow us to observe the thickness of the distribution of the different clusters. The black cluster (cluster 1) takes on a broader range of higher $|z|$ values, while the green cluster (cluster 2) is more centralized at $|z|$ $\approx$ 0. As the green cluster (cluster 2) divides and more clusters are created, the $|z|$ densities stay peaked at $|z|$ $\approx$ 0, with each additional, younger, cluster becoming more localized near the center. 

\begin{figure*}[]
  \centering
  \includegraphics[width=.95\linewidth]{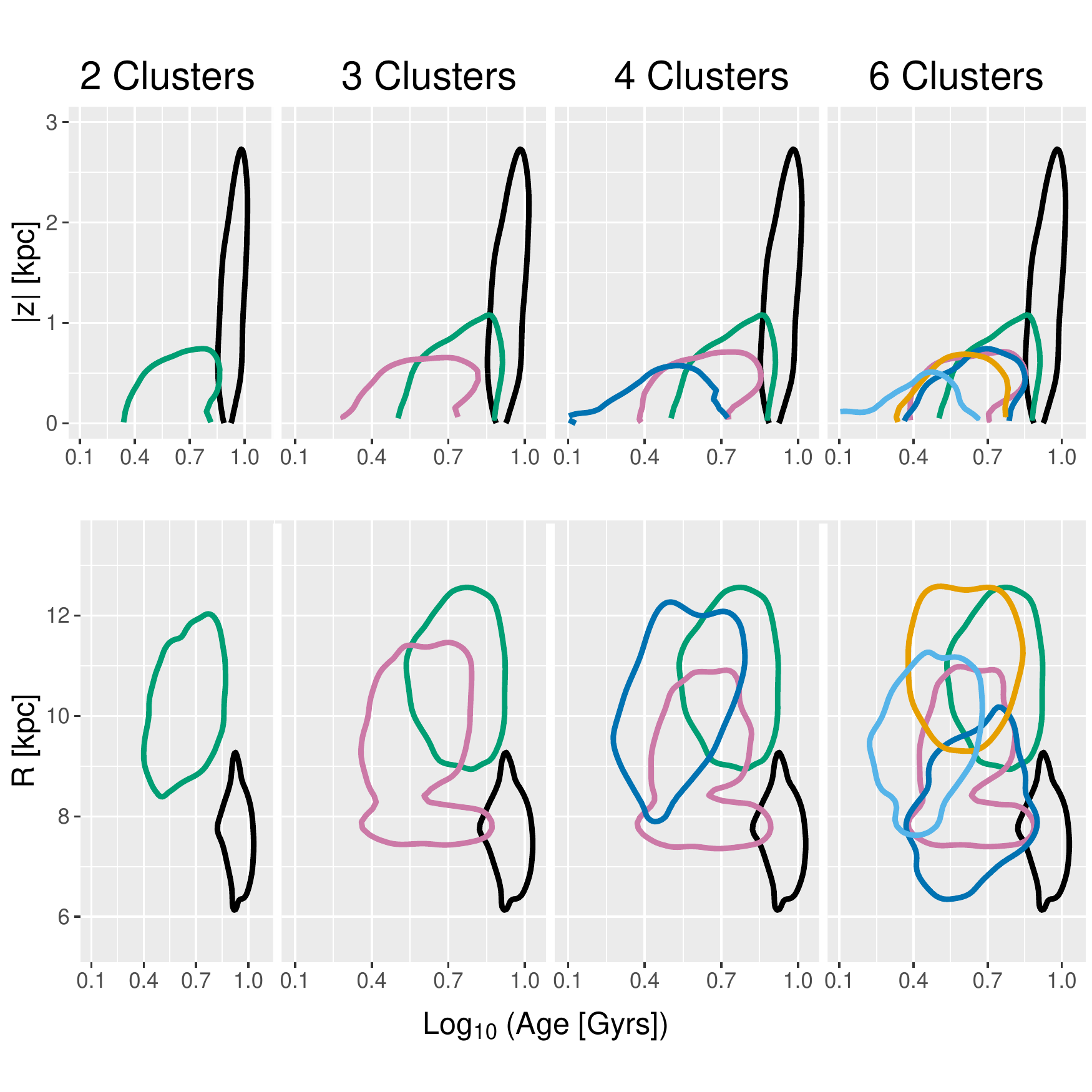}
  \caption{The clusters shown in Figure \ref{fig:All_Clusters}, projected into (\rgal,$|z|$) spatial planes as a function of age. The clusters show different trends, with the centroids shifting in $|z|$-age and living in separate areas in \rgal-age. The contour lines shown contain 50\% of the stars in each cluster.}
  \label{fig:RZage}
\end{figure*}

We further investigate the spatial-age distributions for two through six clusters. Figure \ref{fig:RZage} shows how our groups separate spatially (\rgal\ and $|z|$) as a function of age. The distributions in $|z|$-age show a shift in the density centers, with the cluster centroids showing a positive correlation between $|z|$ and age. As the stellar sample is divided into more clusters, the youngest group of stars consistently gets more localized about the Galactic mid-plane. On the other hand, the primary centers of the clusters separate in no obvious pattern in the \rgal-age plane; the oldest stars are localized to \rgal\ $\approx$ 8 kpc with the next oldest primarily centered at about \rgal\ $\approx$ 11 kpc, then the third oldest cluster is primarily centered at \rgal\ $\approx$ 9 kpc, the fourth oldest more strongly localized near \rgal\ $\approx$ 8 kpc, the fifth oldest spread around \rgal\ $\approx$ 9 kpc, and finally the youngest group of stars located mainly at \rgal\ $\approx$ 9.5 kpc.

\section{Summary and Discussion}
\label{sec:discussion}
In the regime of high-dimensional data, we have the opportunity to go beyond two-dimensional visual classifications of stars into populations, and examine the information in its entirety. In this paper, we use the \apogee\ red clump sample of stars, with stellar ages provided by \cite{Sanders2018} to investigate the distribution  of the data in 19-dimensional abundance-space as well as age and space. Our key findings can be summarized in four points. 

\begin{itemize}
    \item There is coherent (correlated) structure in the abundance values themselves, as seen through their correlation matrix and PCA analysis (see Sections \ref{sec:our_data} and \ref{sec:pca_components}). The correlation structure between standardized abundances shows familial relationships in the light and $\alpha$-elements, which motivates that the data can plausibly be projected into lower dimensions. The first three components of PCA, which describe just over 50\% of the variance of the data, exhibit the same intra-family correlations.     
    \item  Assuming that only two clusters describe the data, the projection of these into the [Mg/Fe]-[Fe/H] plane reveals that there is a subset of stars ($\sim$ 900 stars in this case) that visually appears to be a member of what is typically considered the high-$\alpha$ sequence, but in fact more closely resembles the low-$\alpha$ sequence using our algorithm. That is, clustering by a distance minimization algorithm gives a different two dimensional projection of abundance groups in the [Fe/H]-\alphafe\ plane than that given by a simple visual classification. Clustering in the first two components of both PCA and Isomap confirmed the robustness of this finding, as well as our overall clustering results to lower levels of the hierarchy.            
    \item Our clusters, which are defined in abundance-space alone, separate into populations that are also distinct in both age and spatial distribution. We can decompose our sample into at most six clusters, as this is the highest number of clusters that we can reasonably resolve into different populations in age and (\rgal, $|z|$) distributions given our age errors, which dominate our error budget (a median of $\approx$ 15\% percent). However, the stars can also reasonably be described by a continuum of structure, with no definitive number of clusters comprising our sample.
    \item As we increase the number of clusters, the oldest population (comprised of high-$\alpha$ stars) remains within a single group. This is perhaps indicative of a more coherent chemical evolution event forming these old high-$\alpha$ stars.  
 \end{itemize}

Our first findings on correlations and PCA component structures in the chemical elements confirm the underlying simplicity of abundance space. In agreement with both \cite{Ting2018} and \cite{PJ2018}, we find that the dimensionality of the dataset is much smaller than the full 19 dimensions from our sample. This simplicity can be interpreted as arising from the finite number of nucleosynthetic processes and distinct sites that yield enrichment with distinct abundance patterns \citep[again, in agreement with][]{Ting2018}. We report that the first principal component corresponds to Type II supernova enrichment, the second principal component corresponds to enrichment from Type Ia supernova, and the third principal component captures contributions from dying low mass stars. 

The difference between the distributions of our clusters in the  two-dimensional [$\alpha$/Fe]-[Fe/H] plane and the separation that  assigns stars by-eye into low- and high-$\alpha$ sequences calls into (further) question the extent to which disk decomposition in low-dimensions can be physically interpreted. This difference gives insight into the discordant results when dividing into ``thin'' and ``thick'' disk populations using either space or [$\alpha$/Fe]-[Fe/H].

The clarity of the separation of spatial and age distributions of  our abundance-defined clusters suggests that these higher dimensions add important constraints on Galactic history. For example, when ordered as a sequence in age, the distribution of  our clusters in the [$\alpha$/Fe]-[Fe/H] contradict the simplest pictures of the monotonic chemical evolution in this plane. Rather, our intermediate age and metallicity clusters encompass both the low- and high-$\alpha$ sequences, likely reflective of a complex enrichment history by successive generations of supernovae of both Type II and Type Ia. Indeed, in their hydrodynamical simulations of disk formation \cite{Clarke2019} find  star forming clumps that form on the low-$\alpha$ sequence, migrate to the high-$\alpha$ sequence, and eventually return to the low-$\alpha$ sequence. 

\begin{figure*}
    \centering
    \includegraphics[width=\linewidth]{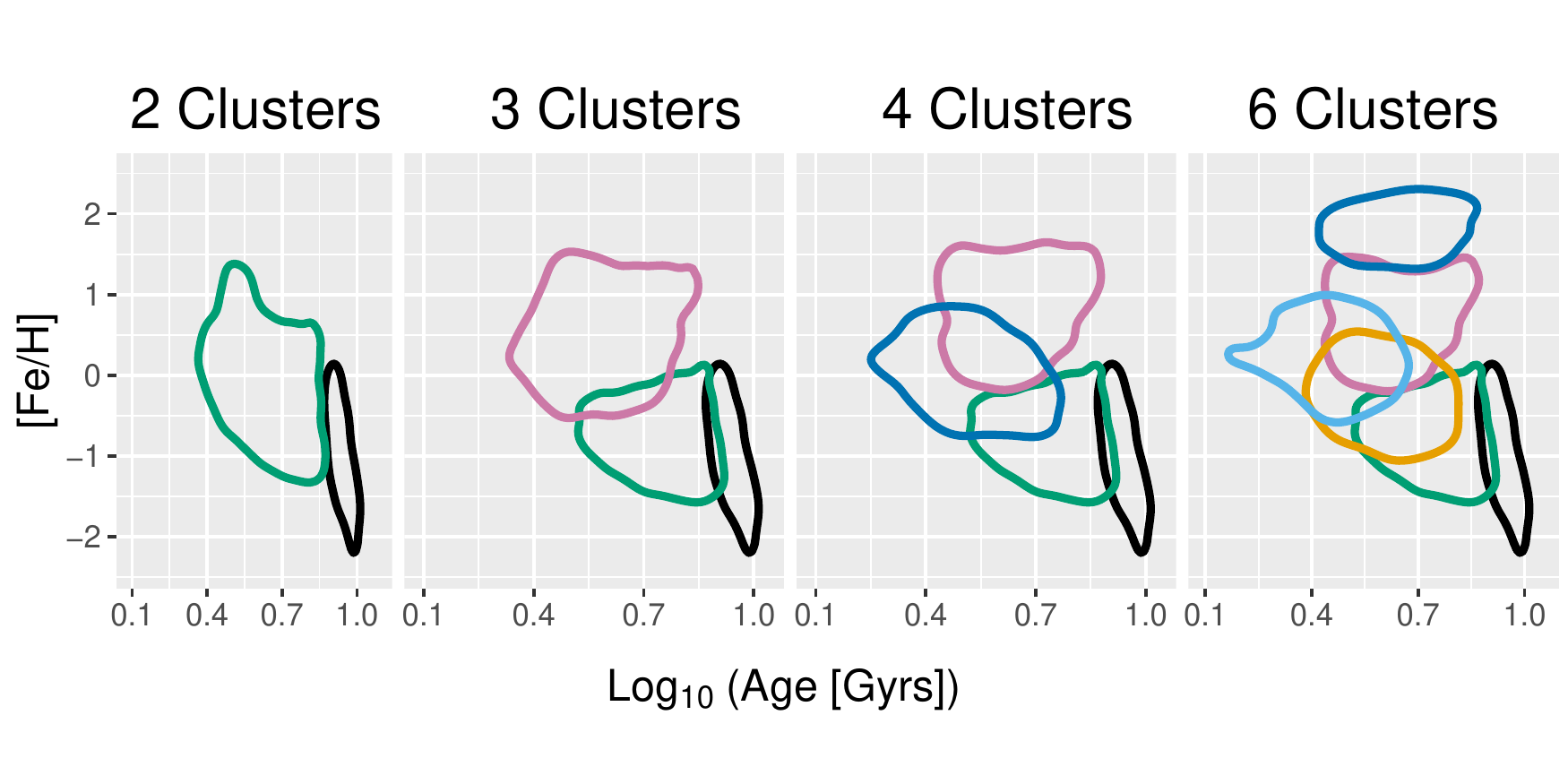}
    \caption{The [Fe/H]-age contours containing 50\% of the stars for each cluster, shown in Figure \ref{fig:RZage}, showing the separation of the clusters in this plane.}.
    \label{fig:ageFe}
\end{figure*}

Overall, our results provide confirmation of the remarkable  simplicity underlying the abundance distributions of stars in the Galactic disk, as has already been reported from different perspectives in other studies. For example, \cite{Weinberg2019} show using \apogee\ data that the abundance trends can be described by two populations across the Galaxy, and \cite{Ness2019} demonstrate that the age and [Fe/H] of stars on the low-$\alpha$ sequence  can be used to predict [X/Fe] for other elements in the data to within 0.02 dex, on average. Our work is consistent with this picture, as apparent in Figure \ref{fig:ageFe}. 

The age catalogue we use has been determined using both spatial and metallicity priors (\cite{Sanders2018}). In the Appendix (Section \ref{sec:appendix_age}), we compare the spatial-age and [Fe/H]-age distributions using two different age catalogues, where ages have been determined using the spectra directly to propagate asteroseismic ages \citep{Pins2018} using data-driven modeling \citep{Sit2019, Ness2019}. We obtain consistent results from all age catalogues, however the clusters show slightly different distributions in these planes.

\section{Conclusion and Future Prospects}
\label{sec:conclusions}

We conclude that clustering in high-dimensional abundance space is informative about the sequence and locations of forming stellar populations. This is a particularly promising variant of chemical tagging, even if we are unable to go as far as reconstructing the individual {\it physical} clusters in which stars are originally born \citep[e.g.][]{Ness2018, Kamdar2019b, Ting2017}. We by no means claim that there are a discrete number of  stellar populations comprising the disk of the Milky Way (i.e. six as shown in Figure \ref{fig:All_Clusters}). Rather, our clusters characterize the underlying continuous evolution in age and metallicity across the disk, analogous to analyses of mono [Fe/H]-age populations \citep[e.g.][]{Bovy2016}.  

This is a pilot study that outlines the potential of current and future data. Several aspects of our approach could be improved. In order to determine the maximum number of clusters with significantly different age populations, we choose to compare the mean ages. The pairwise multiple comparison test design we propose is provisional --- there are other ways to determine discrete aged clusters. Additionally, comparing means is not representative of distributions when the densities are non-normal or spread out. Ideally distributions would be compared, without worry of small fluctuations misleading results.  

In addition, we do not take into account the abundance measurement errors on the \apogee\ data. Therefore all measurements are weighted equally in our clustering algorithm, where as in reality some stars are measured more precisely than others and some elements are systematically measured more precisely than others. Future datasets (or more sophisticated analyses of existing datasets) are likely to give higher precision measurements of abundances, ages, and locations, which will allow cleaner measurements of distributions in these spaces. 

Overall, our results are indicative of the immense promise of future datasets, which could contain even more informative dimensions. One limitation of the \apogee\ DR14 abundance catalogue is that it is restricted to $\alpha$, iron-peak and light abundances, but is missing the channels from the neutron capture processes, which are sensitive to different production mechanisms. The next data release of \apogee, as well as future mission data \citep{Kollmeier2017}, will also contain valuable neutron capture element measurements \citep{Cunha2017, Hasselquist2017} for millions of  disk stars. This will  enable a more complete abundance clustering analysis to inform our picture of the assembly of the disk. These additional elements could reveal additional details or structures that are not apparent in the 19 (highly correlated) abundances that we use, and explain (for example) the events which lead to the formation of the low- and high-$\alpha$ sequences. 

\section{Acknowledgements} 
We acknowledge helpful conversations with the Milky Way Stars group at Columbia University. K.V.J. is supported by NSF grant AST-1715582 and B.S. is supported by NSF grant DMS-1712822.
M.K. Ness is in part supported by a Sloan Foundation fellowship.

 \section{Appendix}
 
 \subsection{Comparison using other age catalogues}
 \label{sec:appendix_age}
 
 \begin{figure*}
    \centering
    \includegraphics[width=.75\linewidth]{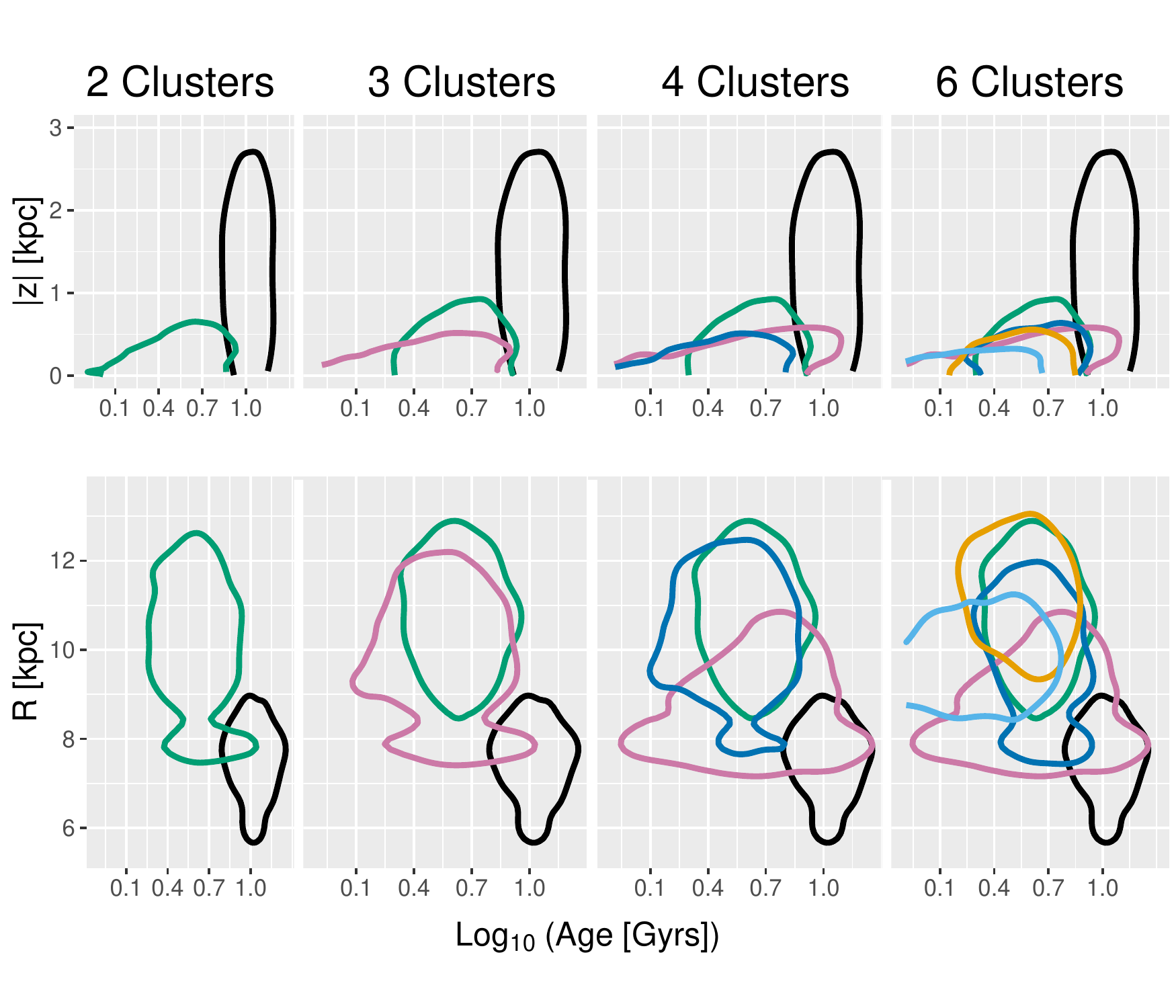}
    \includegraphics[width=.75\linewidth]{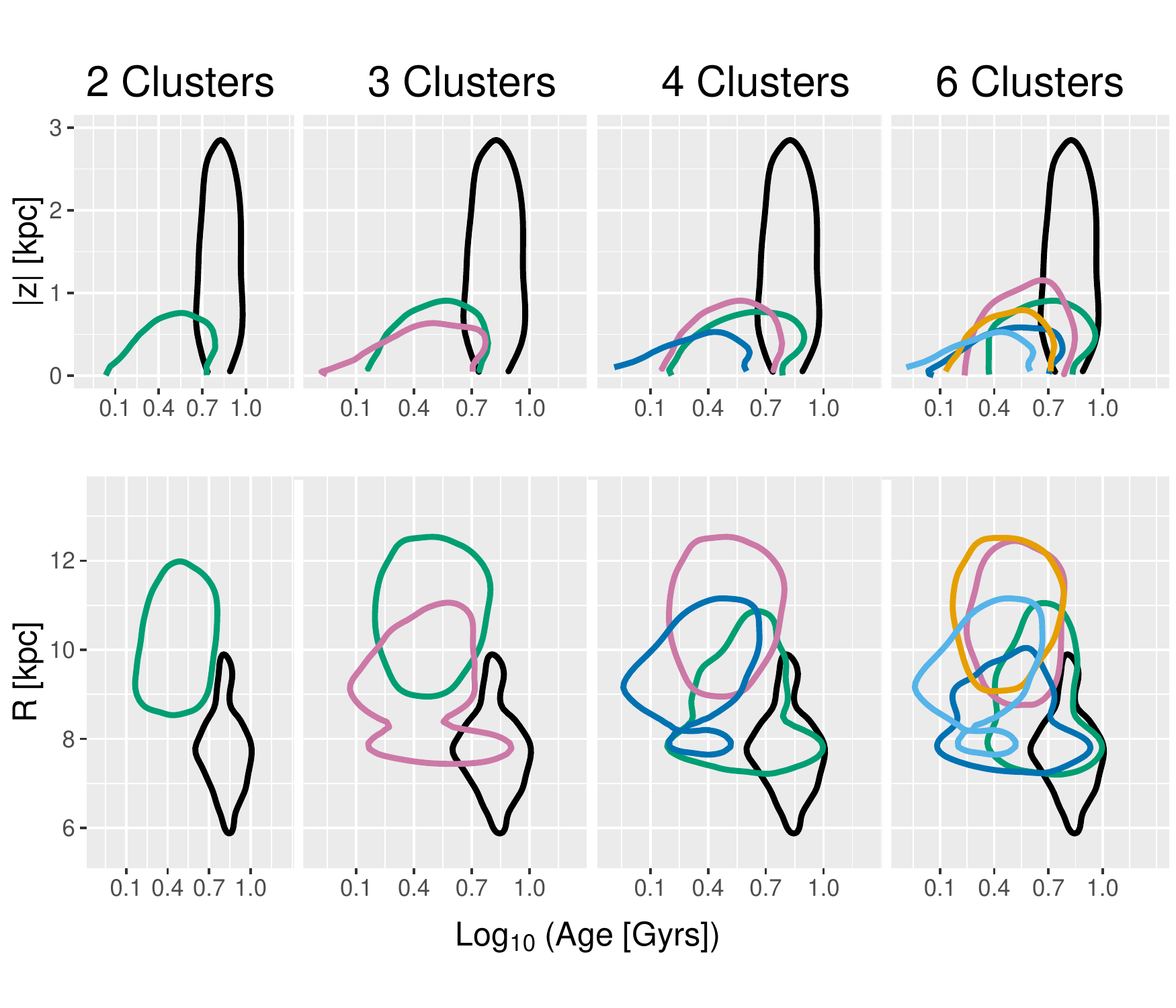}
   
        \caption{The (\rgal,$|z|$)-age planes of the clusters in Figure \ref{fig:RZage}, only using the age catalogue \citet{Ness2016}, at top, and \citet{Sit2019} at bottom.  The clusters show more overlap in the (\rgal,$|z|$)-[Fe/H] planes, but have different means and dispersions.}
    \label{fig:Cannon_RZAge}
\end{figure*}

\begin{figure*}
    \centering
    \includegraphics[width=\linewidth]{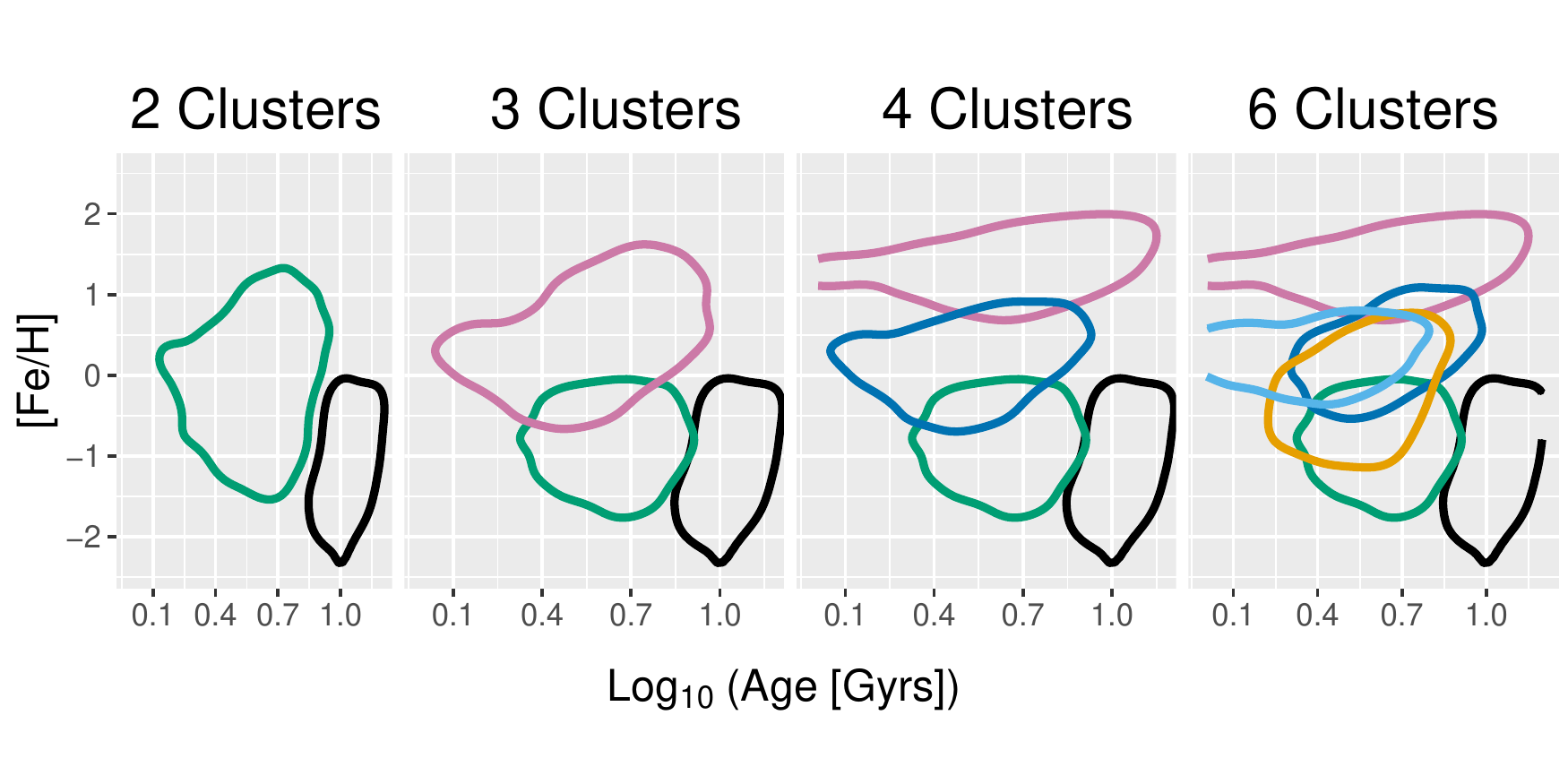}
    \includegraphics[width=\linewidth]{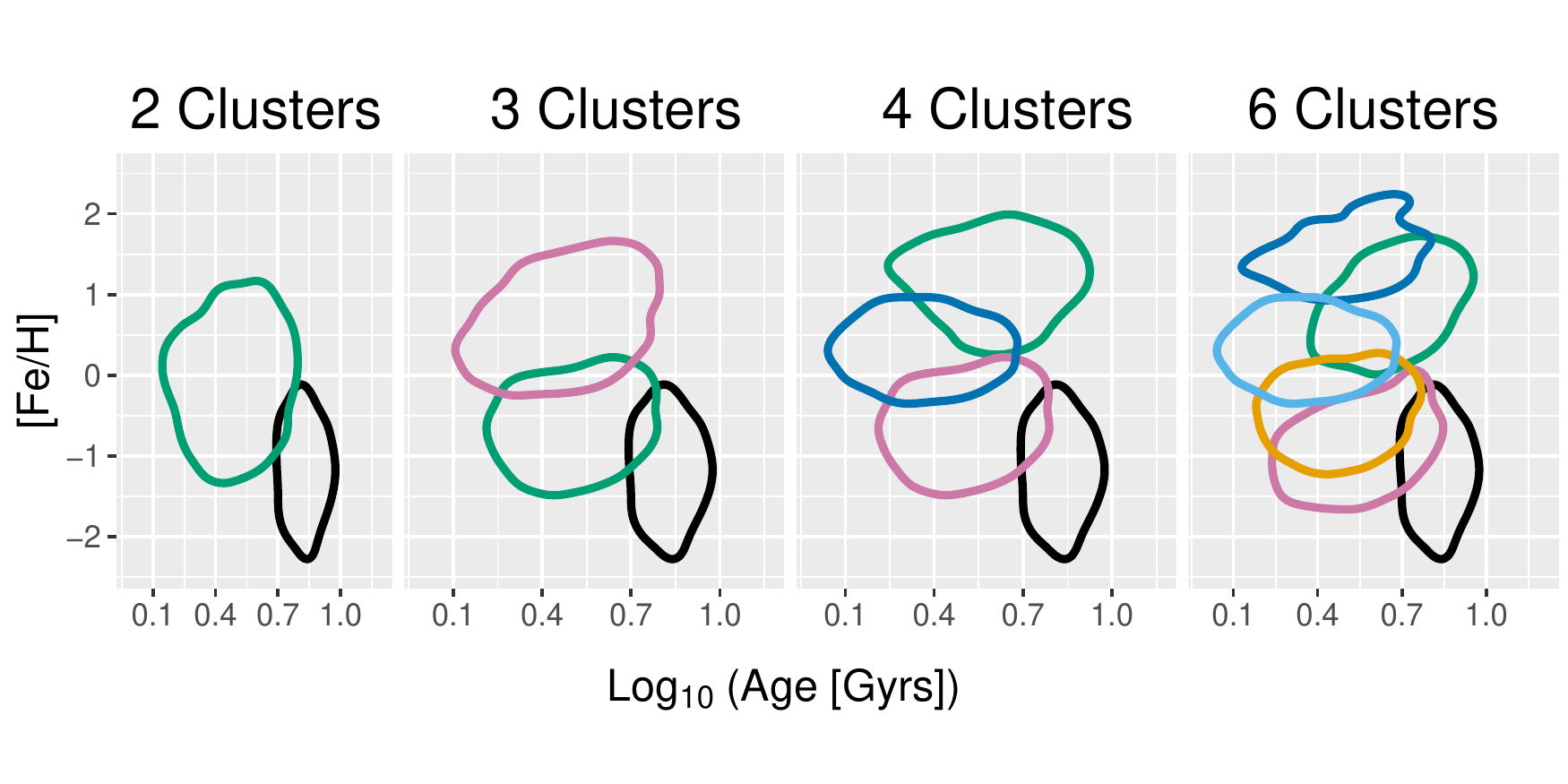}
    \caption{The [Fe/H]-age planes of the clusters in Figure \ref{fig:ageFe}, only using the age catalogue \citet{Ness2016}, at top, and \citet{Sit2019} at bottom. The age errors from these catalogues, at around 40 percent, are about double that of  \citet{Sanders2018}. However, they are derived using the stellar spectra itself and have no prior probability assigned from spatial or abundance information. The different clusters show some overlap in the age-[Fe/H] plane, but have different means and dispersions.}
    \label{fig:Cannon_FeAge}
\end{figure*}
 
 We find that our clusters show separation in planes of [Fe/H]-age and (\rgal,$|z|$)-age. However, we use the age catalogue from \citet{Sanders2018}, which is generated using priors from a model of the spatial distribution of populations in the Galaxy and metallicity priors \citep[see also][]{Das2018}. We want to validate that we are not directly recovering these priors. Therefore, we examine two other age catalogues that have been generated differently to \citet{Sanders2018}. We use two comparison catalogues: one generated using \apogee\ DR12 spectra in \citet{Ness2016} and one generated using \apogee\ DR14 spectra in \citet{Sit2019}. Both approaches employ the data-driven methodology of \tc\ directly on the stellar spectra \citep{Ness2015}. However, there are important differences. The earlier work employs a smaller training set using the then available APOKASC catalogue \citep{P2014}, as well as DR12 spectra and propagates this age information to the remaining DR12 across the parameter range where the survey data is representative of the training data. The later work uses the far larger training set of stars  in \citet{Pins2018} and the DR14 spectra, propagating age to a larger set of stars across a broader range in stellar parameters as well as simultaneously inferring individual abundances. The other major difference between the \citet{Ness2016} and \citet{Sit2019} catalogues is that the latter uses a separate model to infer ages for low and high-$\alpha$ stars, as motivated by \citet{Ness2019}. 
 
There are 17,152 stars in common with our analysis using the \citet{Sanders2018} ages from \citet{Ness2016}, and 25,924 stars in common with \citet{Sit2019}, respectively. Given the different number of stars, we repeat our analysis in its entirety for these other age catalogues. Using the metric described in Section \ref{sec:groups_age}, we find we can go to fewer clusters (in both cases 5 clusters). This is a consequence of the fewer number of stars being analyzed and of the larger age errors (about 40\% for both catalogues). We show the [Fe/H]-age and (\rgal,$|z|$)-age planes for these age catalogues in Figures \ref{fig:Cannon_RZAge} and \ref{fig:Cannon_FeAge} to compare with our results shown in the paper (Figures \ref{fig:RZage} and \ref{fig:ageFe}). The primary difference between these results and that in our main analysis is the less dramatic separation in the spatial planes with age. The discernible trend of younger stars living closer to the Galactic mid-plane found using \cite{Sanders2018} ages is less apparent for both age catalogues, along with the separation in \rgal-age using \citet{Ness2016} ages. However, we still see distinct cluster centers in \rgal-age using \cite{Sit2019} ages for up to four clusters. Additionally, both age catalogues show cluster separation in [Fe/H]-age, with trends reminiscent of the ones found in Figure \ref{fig:ageFe}. Overall, results using ages from the \citet{Ness2016} and \citet{Sit2019} catalogues compliment our main results found.

\subsection{Algorithms and additional figures}

\begin{algorithm*}
\caption{Ward's Minimum Variance Agglomerative Hierarchical Clustering}
\label{alg:Wards}
\begin{algorithmic}[0]
\STATE \textbf{Step 1}: Start with each star as its own cluster.

\STATE \textbf{Step 2}: Create a dissimilarity matrix between cluster $i$ and cluster $j$ for all $i$ and $j$. Let $C_*$ be cluster $*$ and $c_*$ be its cluster center. Then the dissimilarity matrix is:
$$\delta^2(C_i, C_j) = ||c_i-c_j||^2 = \sum_{k=1}^{d=19}(c_i(k)-c_j(k))^2$$
 where 
 $$c_* = \frac{1}{|C_*|}\sum_{x\in C_*}x \in \mathbb{R}^{19}$$
 
 \STATE  \textbf{Step 3}: Combine the least dissimilar clusters $C_i^*$ and $C_j^*$ and update the dissimilarity matrix: 
 $$\delta^2(C_i^* \cup C_j^*, C_k) = \frac{|C_i^*| + |C_k|}{|C_i^*| + |C_j^*| + |C_k|} \delta^2(C_i^*, C_k) + \frac{|C_j^*| + |C_k|}{|C_i^*| + |C_j^*| + |C_k|} \delta^2(C_j^*, C_k)-\frac{|C_k|}{|C_i^*| + |C_j^*| + |C_k|} \delta^2(C_i^*, C_j^*)$$
 
 \STATE \textbf{Step 4}: Repeat Step 3 until all clusters are combined and one large cluster containing all the stars remains.
 
\end{algorithmic}
\end{algorithm*}

\begin{algorithm*}
\caption{Isomap} \label{alg:Isomap}
\begin{algorithmic}
\STATE \textbf{Step 1}: Find the k nearest neighbors for each data point. Construct a neighborhood graph $G$, with the edge length $G_{i,j}$ equal to the euclidean distance between the two neighbors if star $j$ is a neighbor of star $i$.

\STATE \textbf{Step 2}: Create a geodesic distance matrix, $d_G$, that approximates the shortest path between all pairs of points using Dijkstra's algorithm (Algorithm \ref{alg:dijkstra}).

\STATE \textbf{Step 3}: Apply classical Multidimensional Scaling to $d_G$ (Algorithm \ref{alg:mds}).

\end{algorithmic}
\end{algorithm*} 

\begin{algorithm*}
\caption{Dijkstra's Alogrithm}\label{alg:dijkstra}
\begin{algorithmic}

\STATE This is pseudo code to find $d_G(i,j)$, the geodesic distance from star $i$ to star $j$.
\STATE \textbf{Step 1:} Start with all stars marked as unvisited nodes with tentative distance $\infty$ and star $i$ as the current node. 

\STATE \textbf{Step 2:} For each of the current node's unvisited neighbors, compare the neighbors current assigned distance to the tentative distance through the current node. Replace the assigned distance if tentative distance is smaller than assigned distance. 

\STATE \textbf{Step 3:} Once all unvisited neighbors of the current node are considered, mark the current node as visited. 

\STATE \textbf{Step 4:} Continue by selecting the unvisited node with the smallest tentative distance and repeat steps 2-4 until star $j$ has been visited. 

\end{algorithmic}
\end{algorithm*}

\begin{algorithm*}
\caption{Multidimensional Scaling}\label{alg:mds}
\begin{algorithmic}

\STATE MDS takes a distance matrix $d_G$ and returns coordinates in a $p$ dimensional space, where $p$ is the pre-specified desired number of dimensions which we choose to be 2.

\STATE \textbf{Step 1:} Create a squared distance matrix $S$ where $S_{i,j} = d_G(i,j)^2$.

\STATE \textbf{Step 2:} Apply double centering through $B=-\frac{1}{2}JSJ$ where $J = 1-\frac{1}{n}11^{'}$ is the centering matrix.

\STATE \textbf{Step 3:} Calculate the eigenvalues and eigenvectors of $B$. Let $\lambda_1,...,\lambda_p$ be the $p$ largest eigenvalues and $e_1,...,e_p$ be the associated eigenvectors. 

\STATE \textbf{Step 4:} Define $E_p$ as the $n\times p$ matrix with eigenvectors $e_1,...,e_p$ as the columns and $\Lambda_p$ as the diagonal matrix of the eigenvalues from step 3. The new lower dimensional coordinates are $E_p\Lambda_p^{1/2}$. 

\end{algorithmic}
\end{algorithm*}

\begin{figure*}[]
  \centering
  \includegraphics[width=.3\linewidth]{Plots_pdf/FE_H_MG_FE.pdf}
  \includegraphics[width=.3\textwidth]{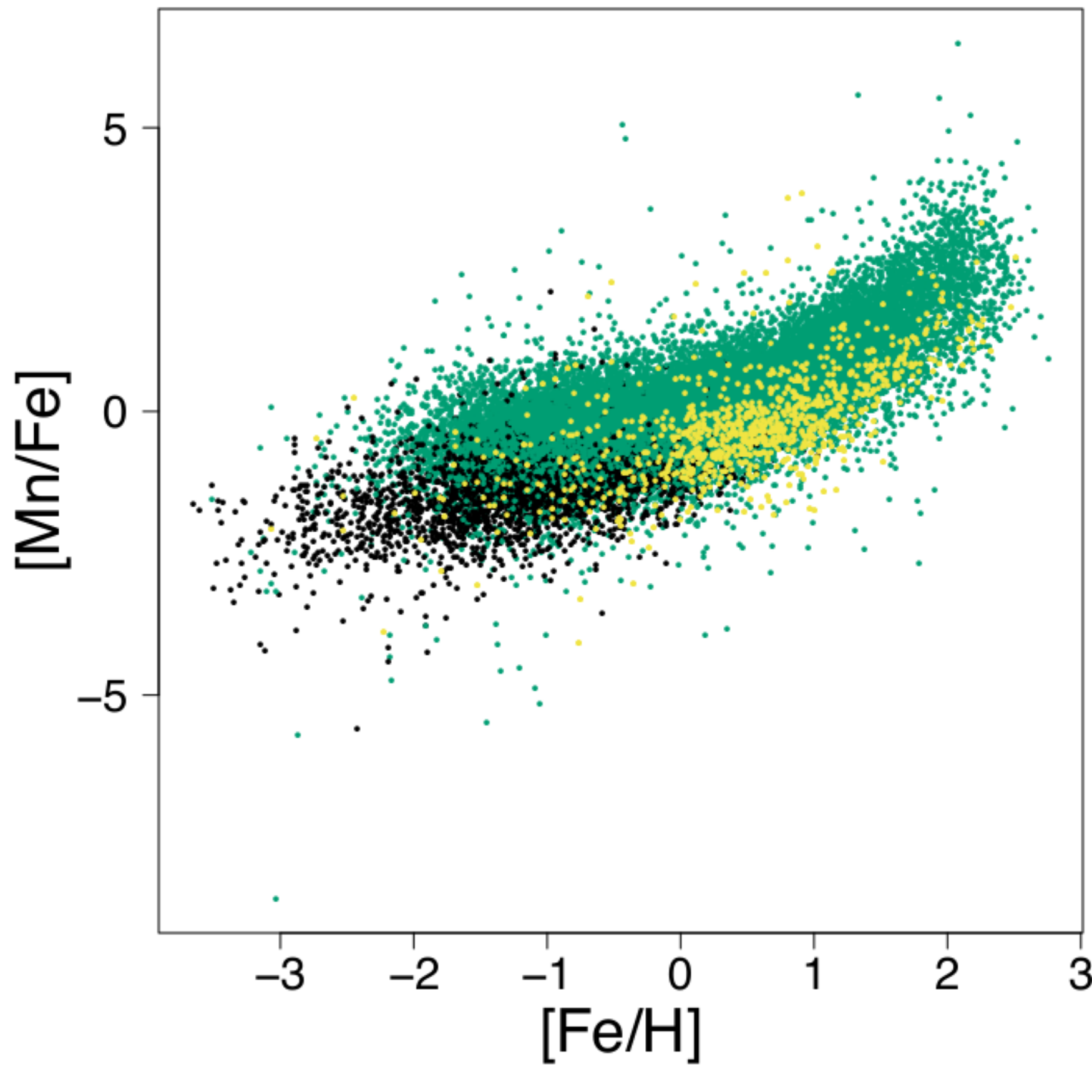}
  \includegraphics[width=.3\textwidth]{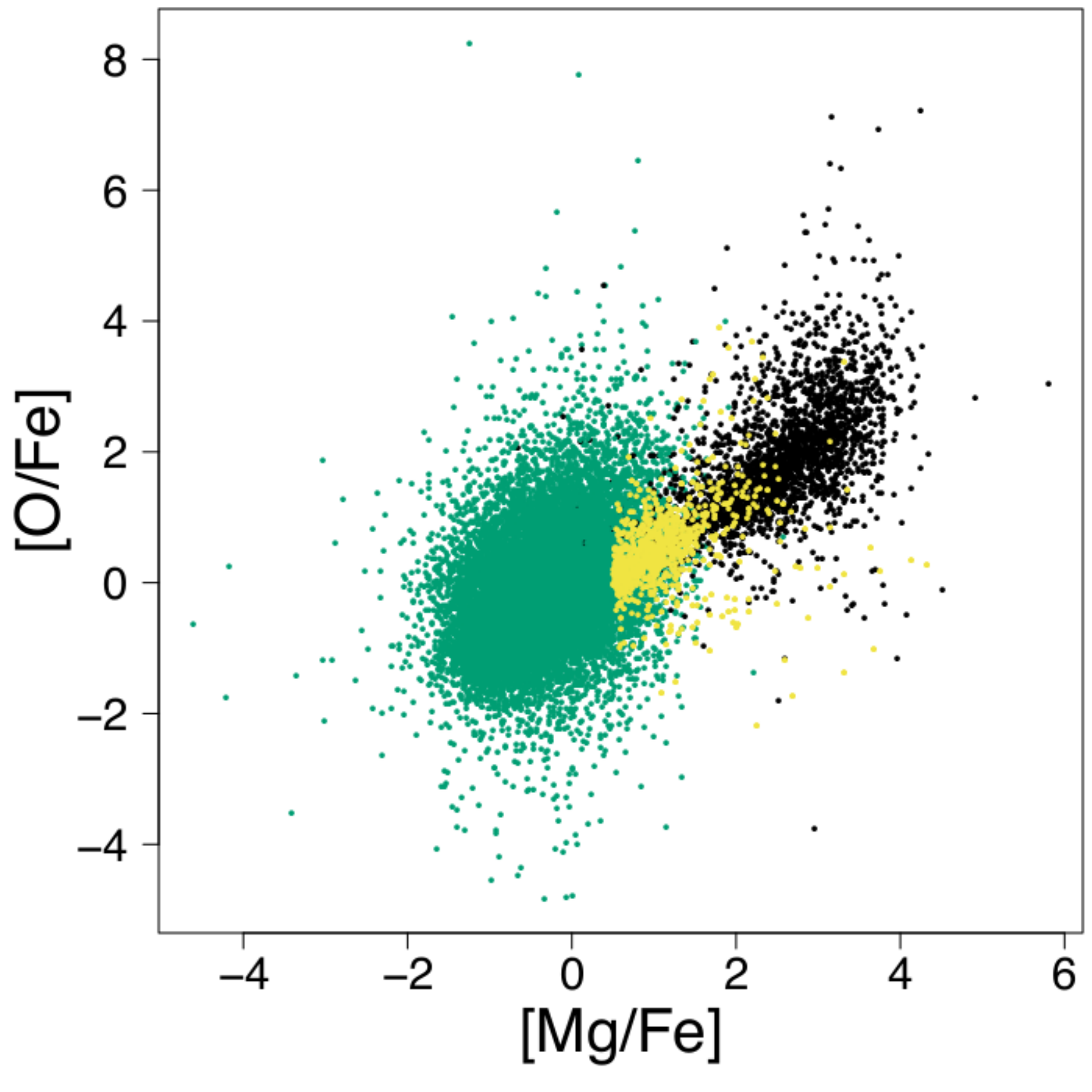}
  \includegraphics[width=.3\textwidth]{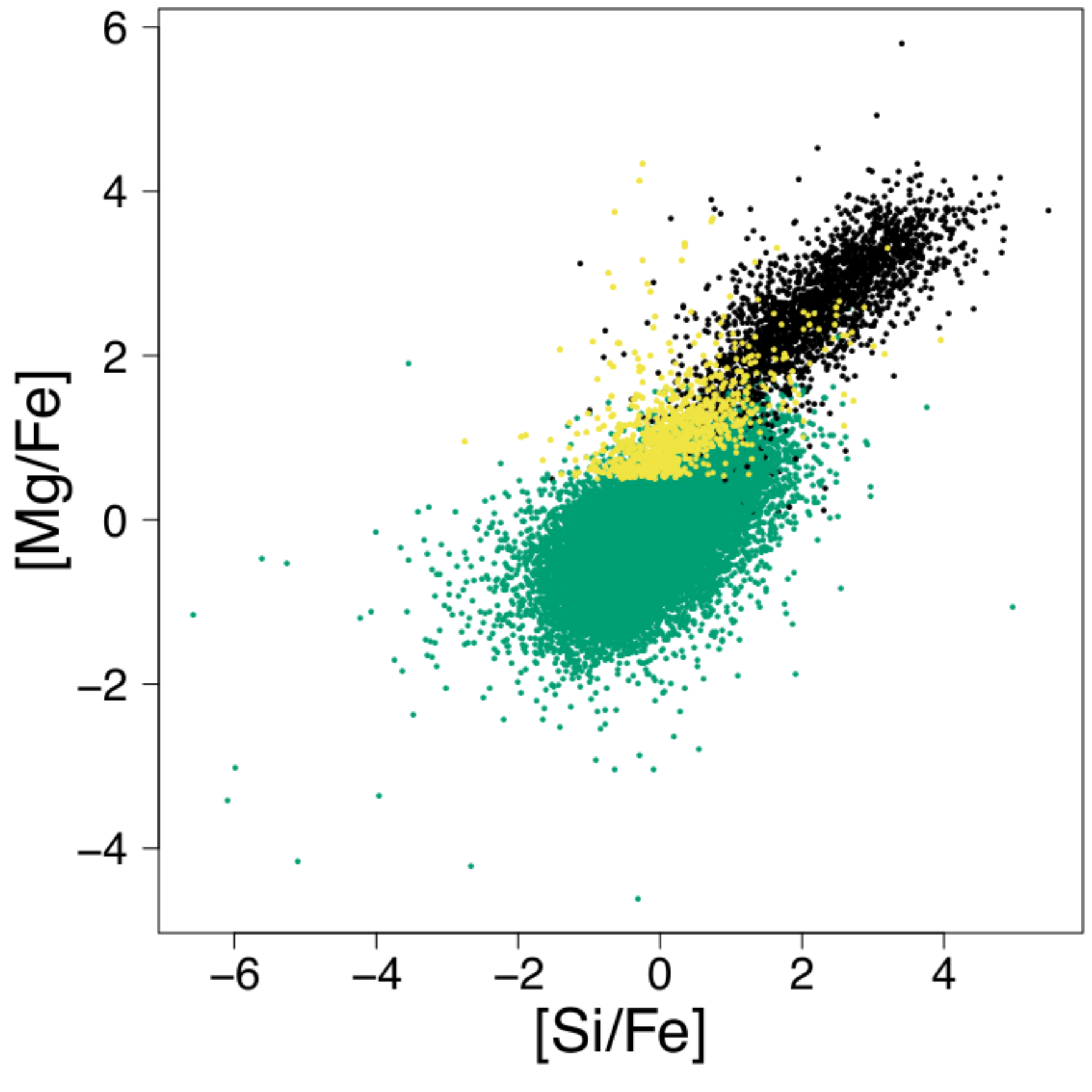}
  \includegraphics[width=.3\textwidth]{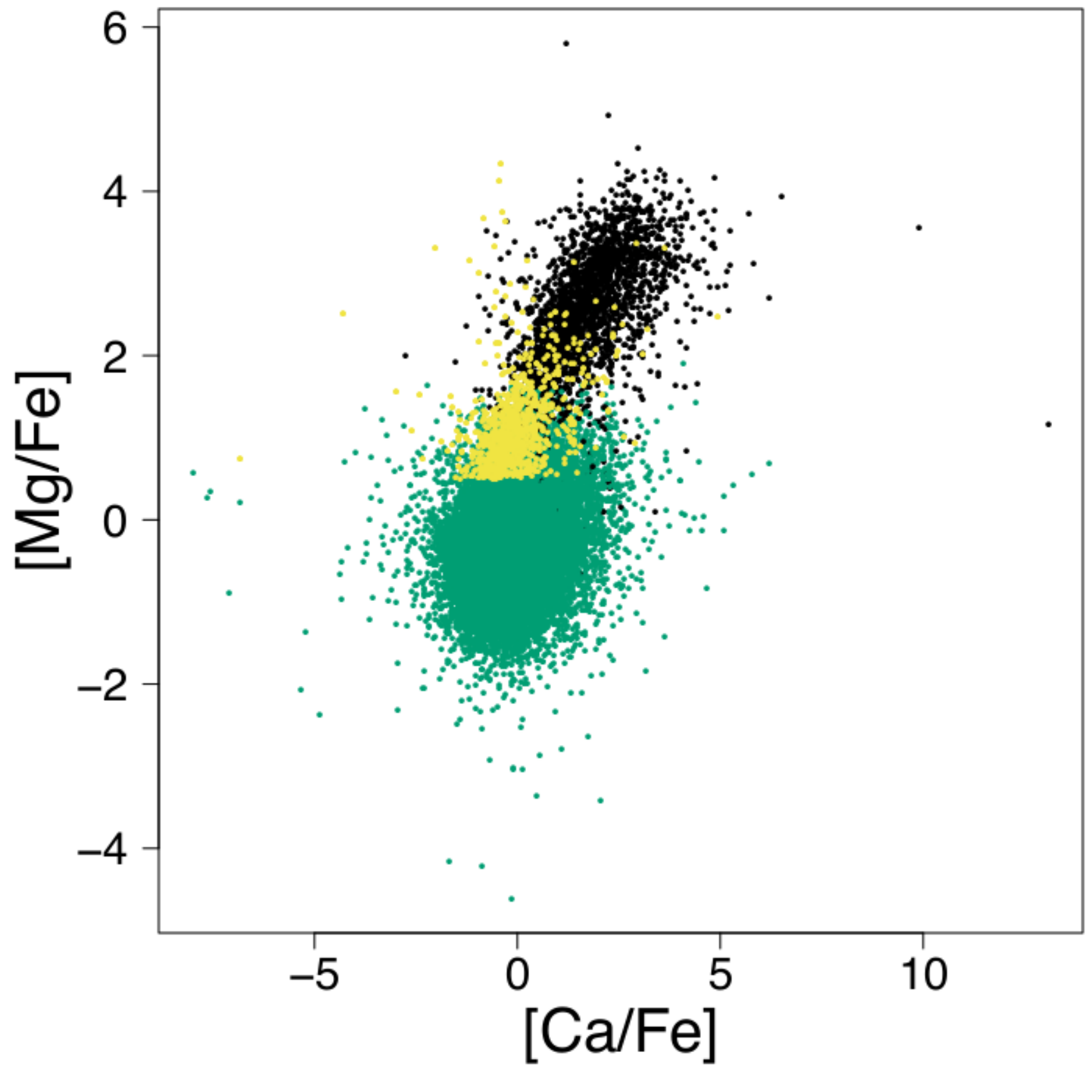}
  \includegraphics[width=.3\textwidth]{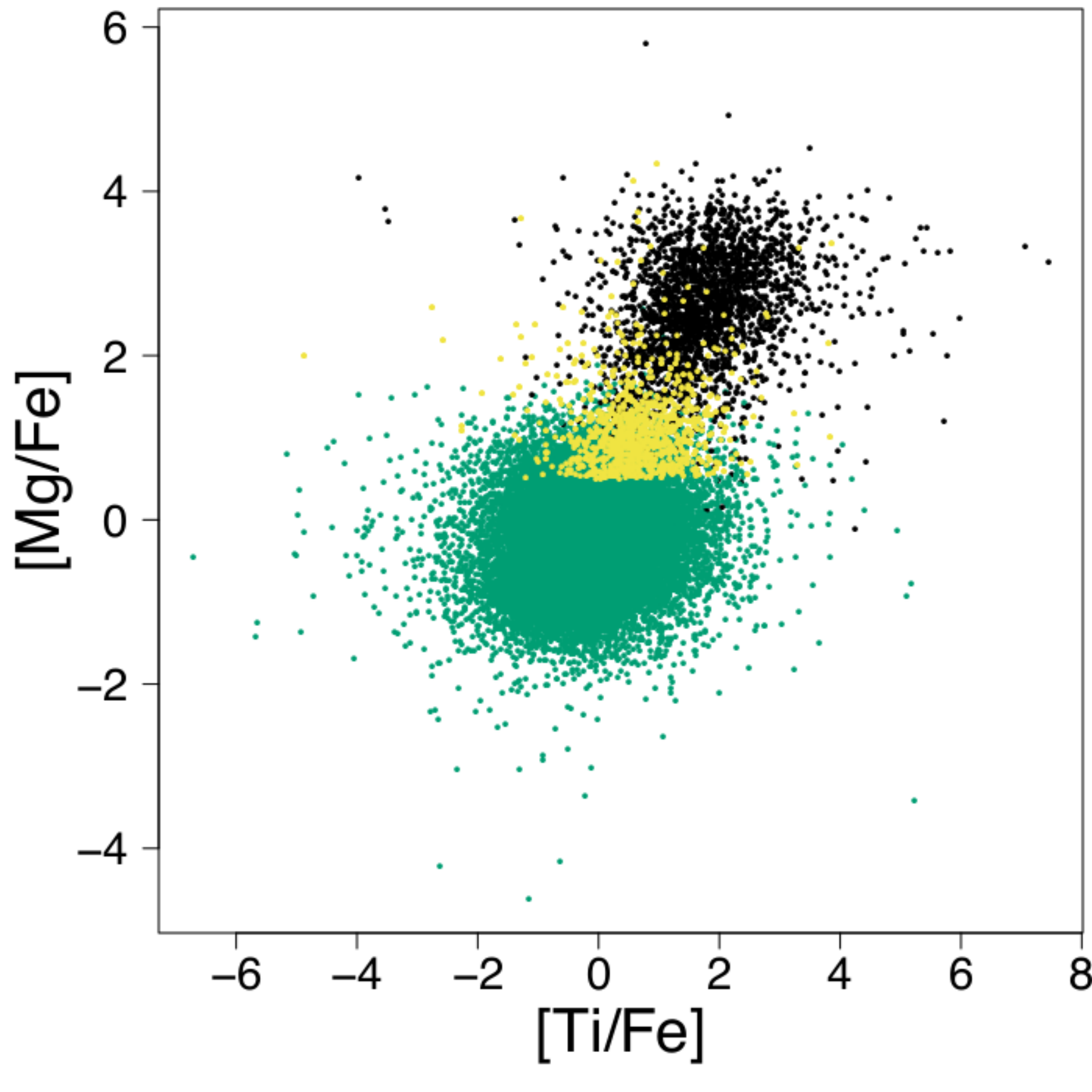}
  \includegraphics[width=.3\textwidth]{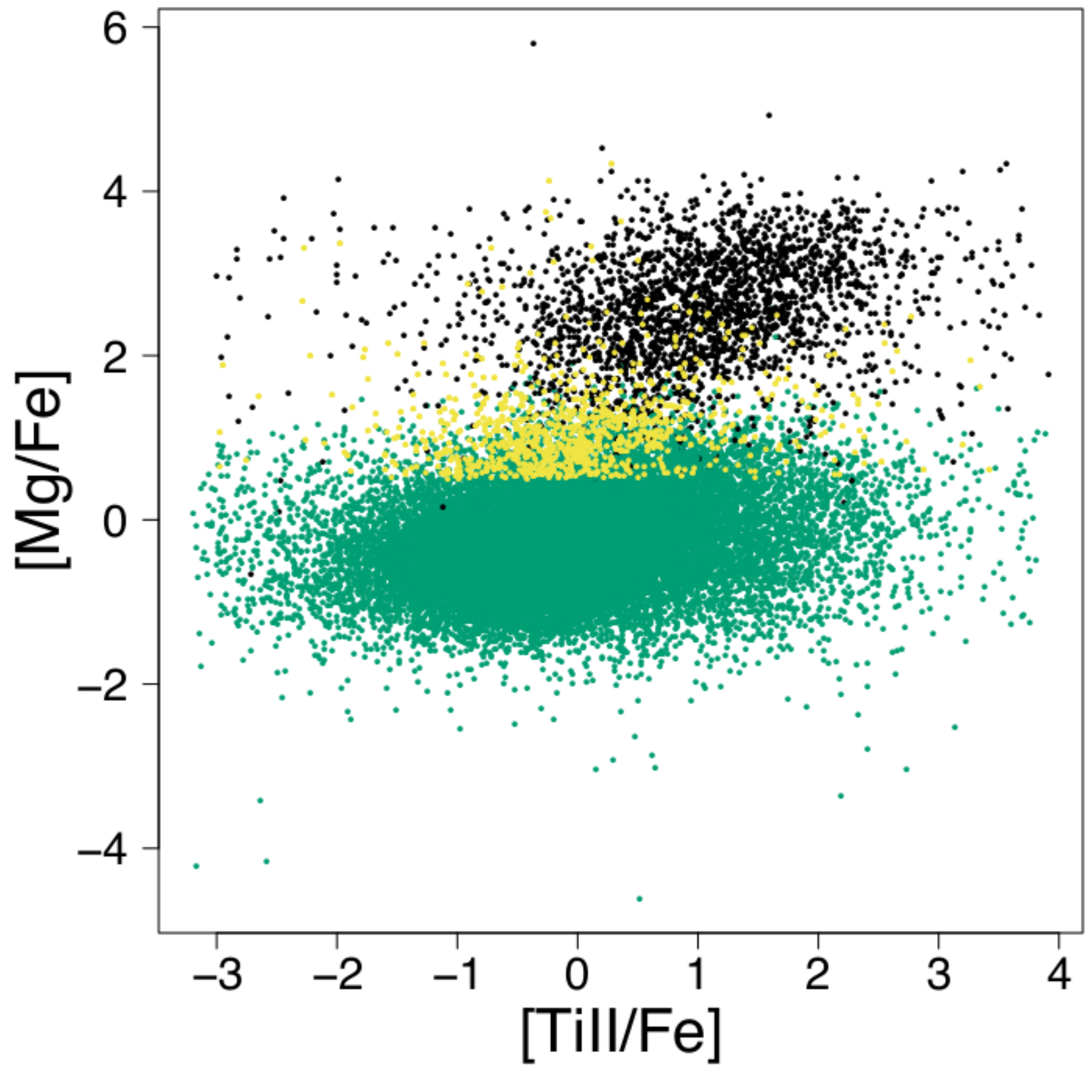}
  \includegraphics[width=.3\textwidth]{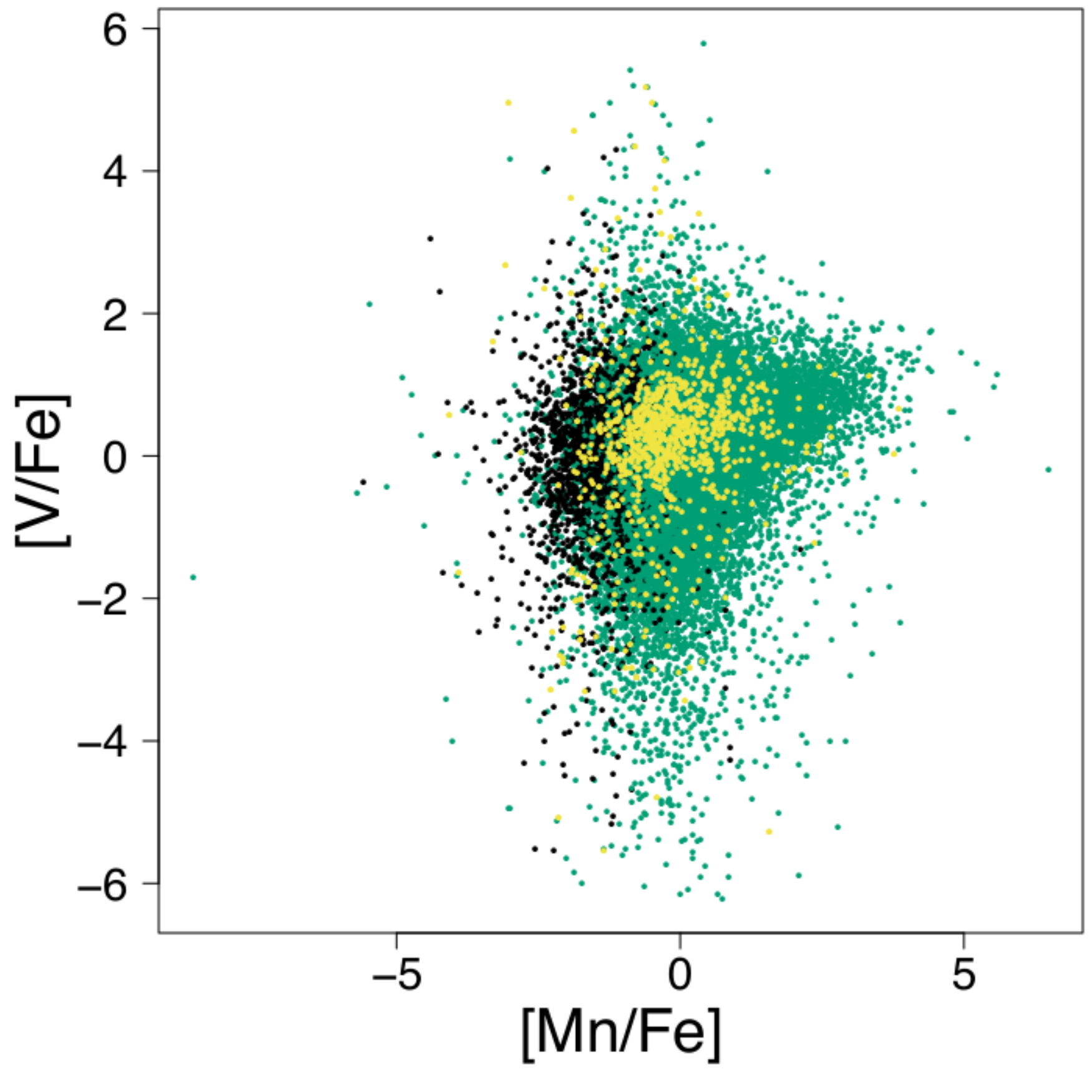}
  \includegraphics[width=.3\textwidth]{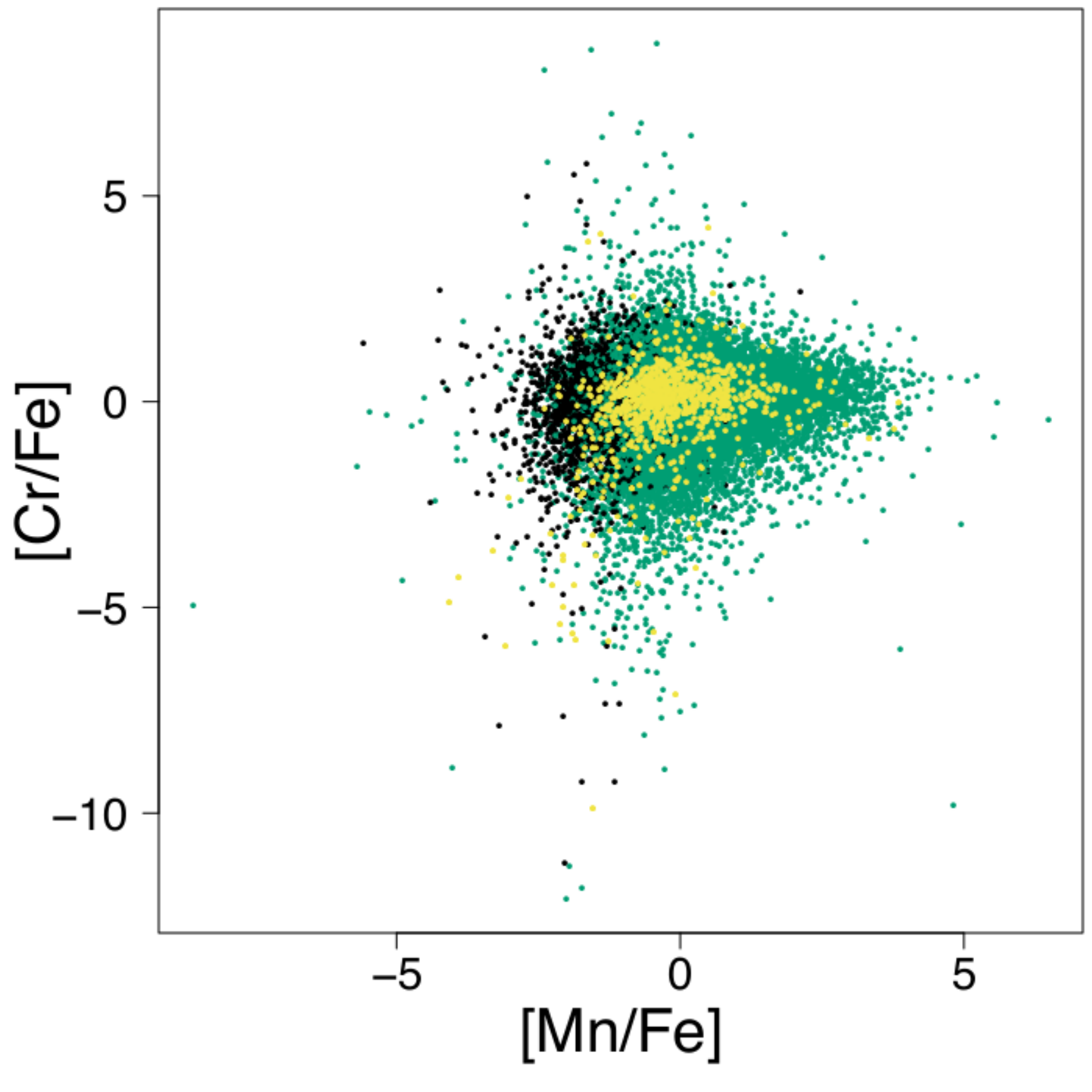}
  \includegraphics[width=.3\textwidth]{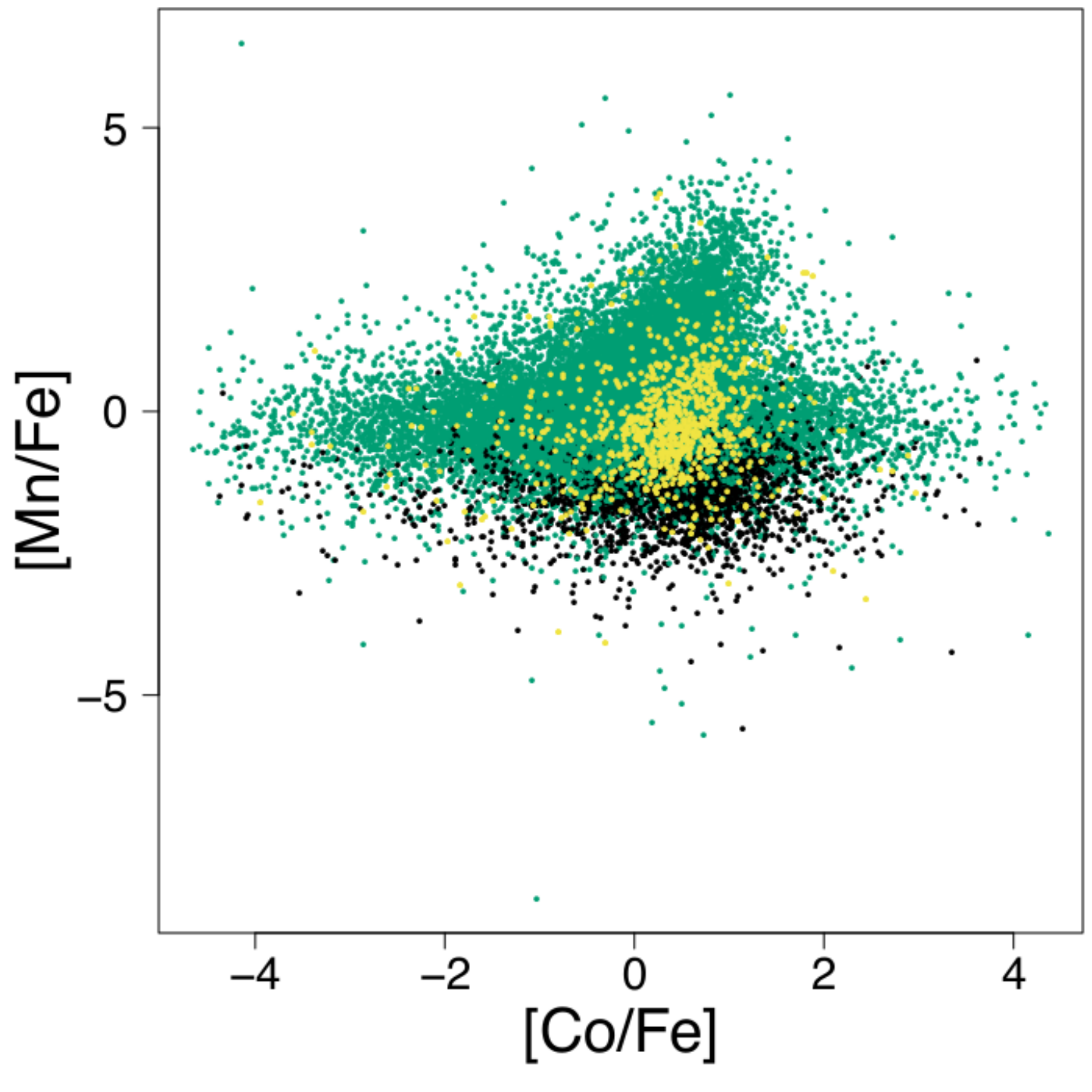}
  \includegraphics[width=.3\textwidth]{Plots_pdf/NI_FE_MN_FE.pdf}
\end{figure*}

\begin{figure*}[]
  \centering
  \includegraphics[width=.3\textwidth]{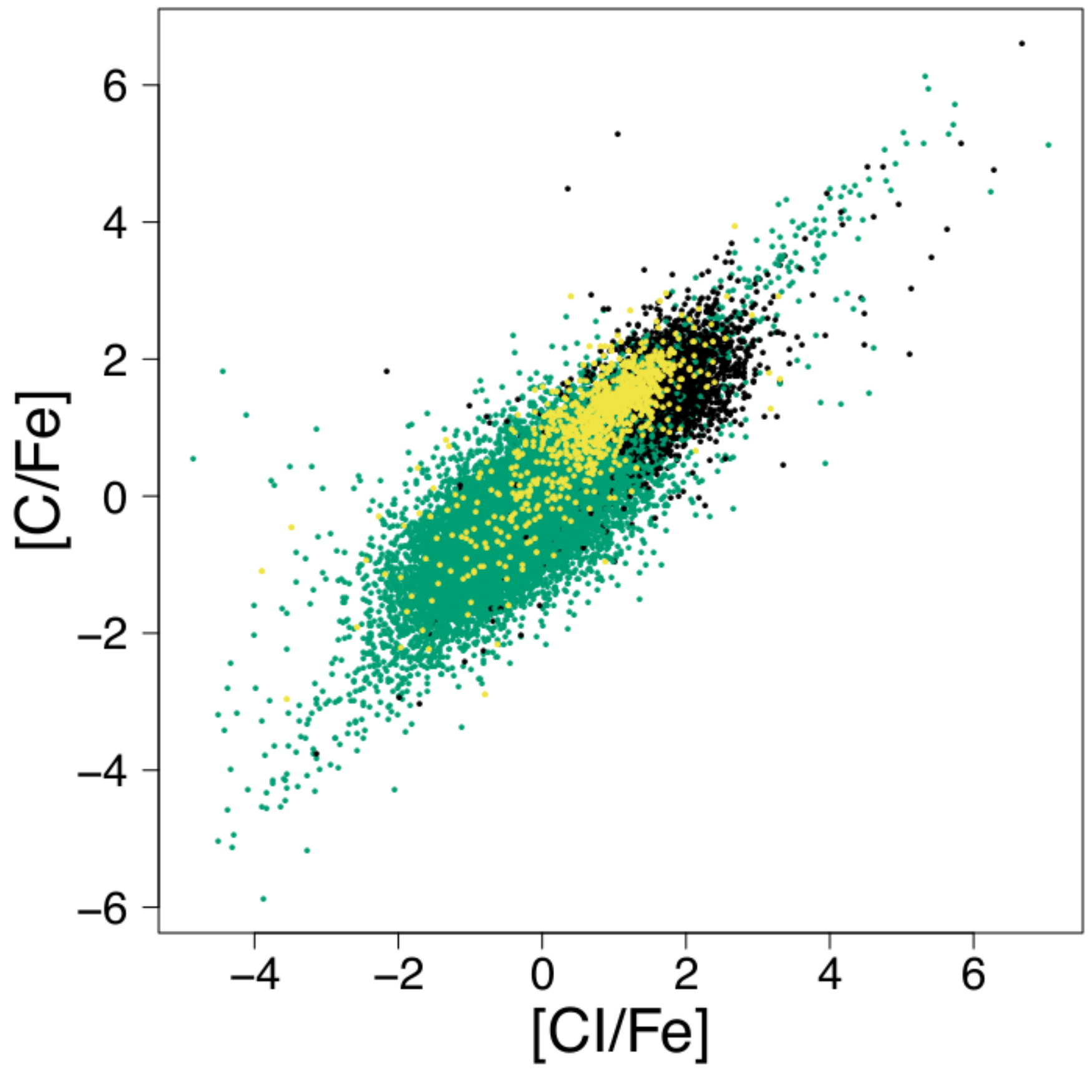}
  \includegraphics[width=.3\textwidth]{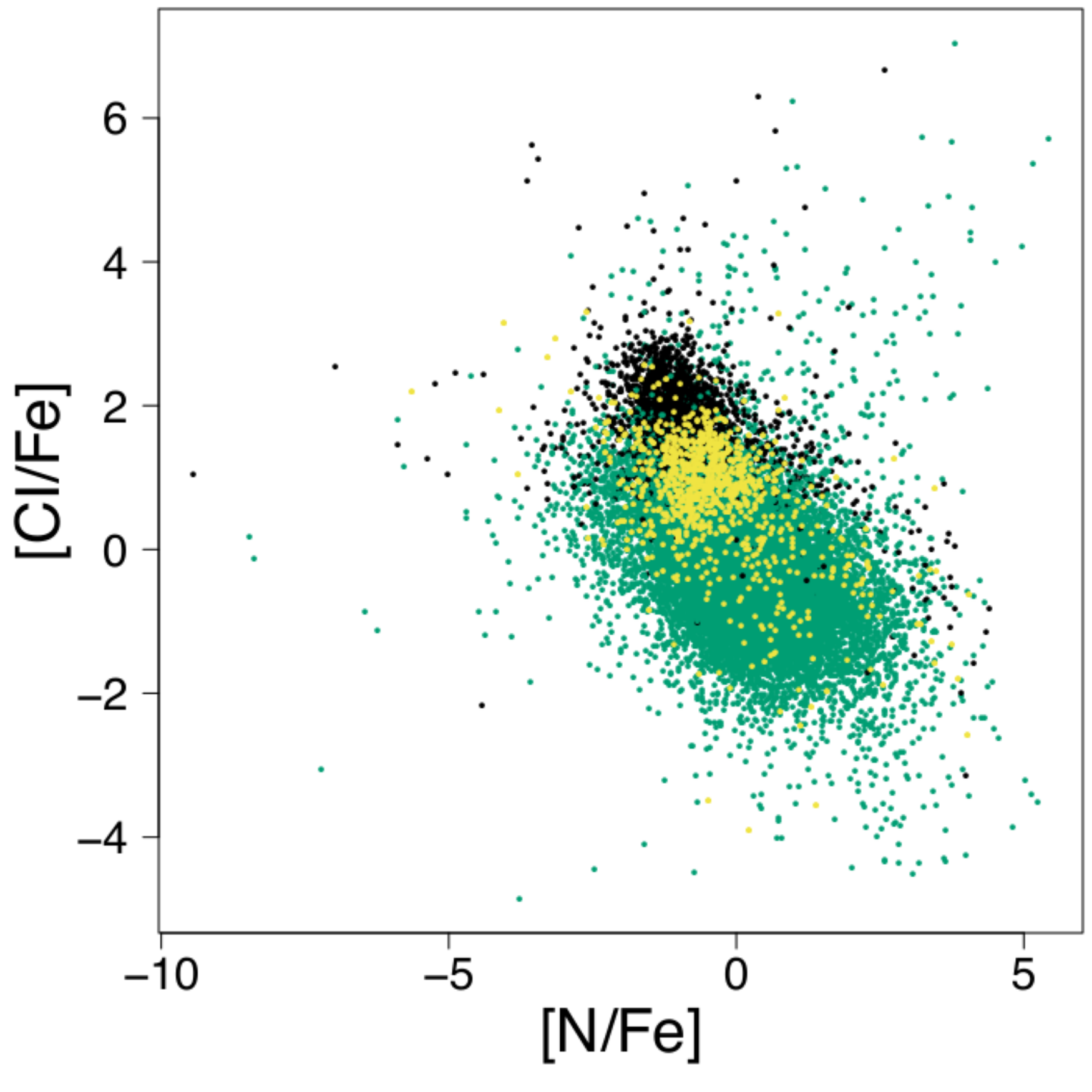}
  \includegraphics[width=.3\textwidth]{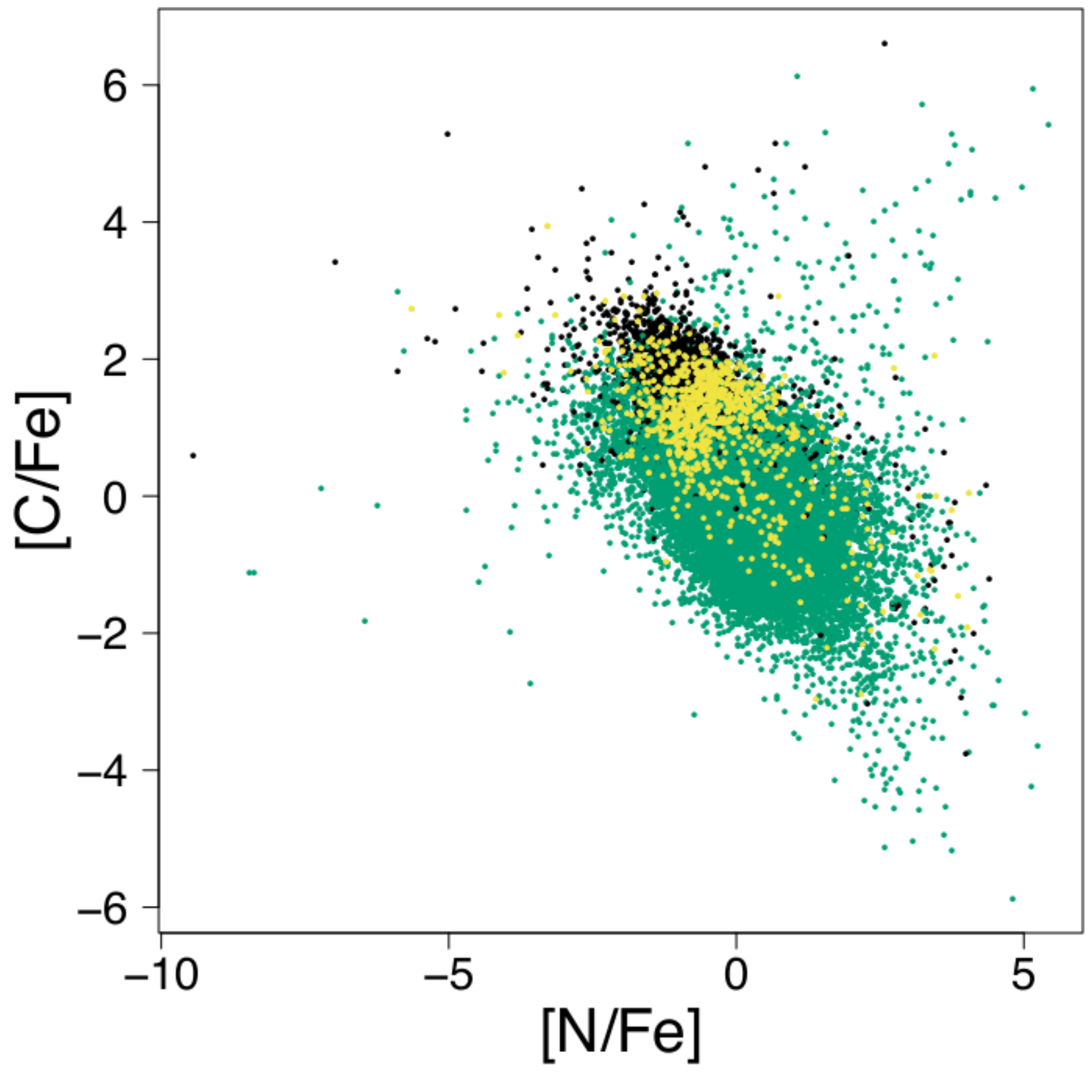}
  \caption{Additional abundance-abundance planes.}
\end{figure*}

\pagebreak
\bibliography{ClusteringAbundances.bbl}
\end{document}